\documentclass[10pt,a4paper]{article}
\usepackage[latin1]{inputenc}
\usepackage{amssymb}
\usepackage{latexsym}
\usepackage{color}
\usepackage{amsmath}
\usepackage{amsfonts}
\usepackage{amsthm}
\usepackage{varioref}
\usepackage{pdfsync}
\usepackage{cancel}
\usepackage{enumitem}

\makeatletter \@addtoreset{equation}{section}

\def \benumlab#1{\begin{enumerate}[label={\rm \bf{(#1{\arabic{*}})}}, ref={\rm #1{\arabic{*}}}]}
\def \enumlab{\end{enumerate}}
\def \benumlabi#1{\begin{enumerate}[label={\rm {(#1\roman{*})}}, ref={\rm{(#1\roman{*})}}]}
\def\abs#1{\left|#1\right|}

\def \Int{\displaystyle\int}

\def \Liminf{\displaystyle\liminf}
\def \Limsup{\displaystyle\limsup}

\def \E{\mathbb{E}}
\def \F{\mathbb{F}}

\def \R{\mathbb{R}}

\def\P{\mathbb{P}}
\def\Q{\mathbb{Q}}

\def\Cc{{\cal C}}

\def\Fc{{\cal F}}

\def\Hc{{\cal H}}

\def\Jc{{\cal J}}

\def\Lc{{\cal L}}

\def\Oc{{\cal O}}

\def\Qc{{\cal Q}}

\def\Uc{{\cal U}}

\def\ep{\hbox{ }\hfill$\Box$}
\def\reff#1{{\rm(\ref{#1})}}
\def\be{\begin{eqnarray}}
\def\ee{\end{eqnarray}}
\def\bal{\begin{aligned}}
\def\eal{\end{aligned}}
\def\beq{\begin{equation}}
\def\eeq{\end{equation}}
\def\beqq{\begin{equation*}}
\def\eeqq{\end{equation*}}
\def\b*{\begin{eqnarray*}}
\def\e*{\end{eqnarray*}}

\def\x{\times}

\def\={\;=\;}

\def\pourtout{\mbox{ for all } }

\def\.{\;.}

\def\eps{\epsilon}

\def\vp{\varphi}
\def\1{\mathbf{1}}

\def\Esp#1{\mathbb{E}\left[#1\right]}

\def\Pro#1{\mathbb{P}\left[#1\right]}

\def\Tr#1{\mbox{\rm Tr}\left[#1\right]}

\theoremstyle{plain}
\newtheorem{theorem}{Theorem}[section]
\newtheorem{proposition}[theorem]{Proposition}

\newtheorem{assumption}[theorem]{Assumption}

\newtheorem{lemma}[theorem]{Lemma}
\theoremstyle{definition}
\newtheorem{remark}[theorem]{Remark}

\def\a{a\hspace{-0.65mm}\iota}

\def\st{\text{ s.t. }}
\def\lpt#1{\\[#1pt]}

\labelformat{assumptionCU}{(${\underline{{\rm\bf C}}}$)}
\labelformat{assumptionCV}{(${{{\rm\bf C}}}$)}

\def\step#1#2{\noindent \emph{Step #1:~#2}}

\def\vc{\vartheta}

\definecolor{darkred}{rgb}{0.8,0,0}
\definecolor{darkblue}{rgb}{0,0,0.7}
\definecolor{darkgreen}{rgb}{0,0.4,0}

\def\Lmf{{\mathfrak L}}
\def\Amf{{\mathfrak A}}
\def\D{{\mathfrak D}}
\def\Di{{\mathfrak{D}_{<T }}}
\def\DT{{\mathfrak{D}_{T}}}

 \def\LSX{\Lc_{\mbox{\tiny SX}}}
 \def\LPlSX {\Lc_{\mbox{\tiny P$|$SX}}}
 \def\LPlSXh {\hat{\Lc}_{\mbox{\tiny P$|$SX}}}
 \def\LStheta{\Lc_{\mbox{\tiny S$\theta$}}}
 \def\LPlStheta {\Lc_{\mbox{\tiny P$|$S$\theta$}}}
 \def\LPlSthetah {\hat{\Lc}_{\mbox{\tiny P$|$S$\theta$}}}
 
 \def\LS{\Lc_{\mbox{\tiny S}}}
  
  \def\LXlSP{\Lc_{\mbox{\tiny X$|$SP}}}
  \def\xib{\xi\hspace{-1.7mm}\xi}
  \def\xibu{\xi\hspace{-1.4mm}\xi}

\def\vs#1{\vspace{#1mm}}

\begin{document}

\title{Hedging under an expected loss constraint\\
with small transaction costs
 \thanks{Research of the first author is
 partly supported by ANR Liquirisk and Labex ECODEC.
 Research of the last two authors was partly supported by the
European Research Council under the grant 228053-FiRM,
 by the ETH Foundation and by the Swiss Finance Institute.
}}

\author{
     Bruno Bouchard\thanks{CEREMADE, Universit\'e
     Paris Dauphine and CREST-ENSAE, bouchard@ceremade.dauphine.fr.}
     \and
     Ludovic Moreau
     \thanks{Department of Mathematics, ETH Zurich , ludovic.moreau@math.ethz.ch.}
    \and
    H.Mete  Soner \thanks{
 Department of Mathematics, ETH Zurich \&
 Swiss Finance Institute, hmsoner@ethz.ch}
 }

\maketitle 

\begin{abstract}
We consider the problem of
option hedging in a market with proportional
transaction costs.  Since  super-replication
is very costly in such  markets, we replace
perfect hedging
with an expected loss constraint.
Asymptotic analysis for small
transaction costs is used
to obtain a tractable model.
A general expansion theory is
developed using the dynamic
programming approach.
Explicit formulae are   obtained
in the special cases of exponential and power utility functions. As a corollary, we retrieve the asymptotics for the exponential utility indifference price.
 \end{abstract}

\vspace{3em}


\vspace{1em}




\section{Introduction}

As well known, in a complete
market with no frictions,
every contingent claim can be replicated
by continuous trading of the underlying asset.
These replicating strategies however typically yield
portfolio processes that are of unbounded variations.
Hence, any size of  transaction cost  renders
this portfolio to have an infinite trading cost.
Indeed, it has been shown that,
generically,  the cheapest super-replicating portolio
is the simple buy and hold strategy leading to a prohibitive cost
 \cite{SonerShreveCvitanic95,LeventalSkorohod97,BoTo00,CvPh99,DeKaVa02,KaLa02,KoPhTo99a,KoPhTo99b}.

Theoretically almost sure replication
is an appealing concept which has
been extensively studied
in the literature.
Firstly, it provides the initial building block
for the utility maximization problems
by providing the exact description of the
wealth processes that enter into the
maximization.  Also it
provides  complete risk aversion
agreeing with all other approaches
and in incomplete markets it yields
the pricing intervals.  When this interval is tight,
it can also have practical uses.
However, since this is not the case in
markets with transaction costs,   one has to
consider instead expected loss criteria
related to the risk attitude of the investors.

In the frictionless
Black-Scholes market
F\"ollmer and Leukert \cite{FoLe99,FoLe00}
studied the quantile and expected shortfall
by exploiting the deep connection
to the Neyman-Pearson lemma, which applies to general complete markets.
A more general  approach for Markovian settings was then
developed in
\cite{bet09,BoDa10,moreau10,BMN12}
for diverse markets including jumps
and several loss criteria.
A particular application of this approach is
the utility indifference
as introduced by Hodges and Neuberger
\cite{HoNe89} in which
the hedging constraint is given
through the maximum utility that
one may achieve without the liability.
However, in the general
formulation of hedging with expected
loss, one can place more than one
constraint \cite{BoVu12} and
consider markets with
general dynamics as well as frictions.

In this paper, we follow
the problem formulation of
\cite{bet09}
and develop a
coherent asymptotic
theory for hedging
problems under an expected
loss criterion,
when the transaction cost is small.
Asymptotic analysis allows for
more tractable formulae.
Our methodology is robust enough
to treat models with general
dynamics and many loss criteria.
For modeling the financial market,
we follow the seminal papers
\cite{CoMa76,Constantinides86}
and the rigorous mathematical
approaches of  \cite{DaNo90,DuLu91,SoSh94}.
For further information
on  utility maximization under transaction costs,
we refer the
reader to the book \cite{KaSa09}
and the references therein.

On the technical side, we build upon the similar theory
that was developed
in the case of  the classical utility maximization.
For this problem,
an extensive theory is now available
starting with the appendix of \cite{SoSh94}.
There are now many rigorous results
 \cite{AMS13,BaSo98,Bichuch12,BiSh13,GeMuSc12,JaSh04,MMS13,PoSoTo11,SoTo12}
 as well as interesting formal derivations
\cite{AtMo04,GoOs10,WhWi97}.
The partial differential
equation (PDE) technique that we use
has its origins in a recent paper \cite{SoTo12}.
It is based on the theory of the viscosity approach
to homogenization of Evans \cite{Evans92}.
This methodology allows for a flexible
asymptotic theory
that applies to markets with multiple
assets \cite{PoSoTo11},
fixed transaction costs \cite{AMS13}
and market impact in factor models \cite{MMS13}. A related  asymptotic analysis is carried out for stochastic volatility models with different time scales  \cite{FPS00,FPSS13}, and for utility maximization asymptotics  \cite{FSZ13}. They also use viscosity solution tools, but their methodology is different.

The asymptotic expansion is derived directly using the
PDE characterization of the expected loss based price.
This equation follows from the
stochastic target formulation with controlled expected loss
as in  \cite{bet09}.  In the frictionless case,
the problem described in subsection \ref{sec: frictionless model} is
 $$
            \pi(t,s,p) := \inf \left\{
                    z\in\R : \Esp{\Psi\left( Z^{t,s,z,\vc}_{T} - g(S^{t,s}_{T}) \right)}\ge p
                    \mbox{ for some } \vc\in \Uc(t,s,z)
                \right\},
$$
where $\Psi$ is the given expected loss function,
$p$ is the given desired threshold,
$g$ is option pay-off,
 $\Uc(t,s,z)$ is the set of admissible controls and
the process $Z^{t,s,z,\vc}$ is the value of the portfolio
with initial stock value $s$, initial wealth value $z$
and control process $\vc$.
The diffusion type dynamics of $Z^{t,s,z,\vc}$
and the exact description of the admissible class
$\Uc(t,s,z)$ are given in section
\ref{sec: models} below.  Then, with the help
of the martingale representation, \cite{bet09} converts this
problem into a standard stochastic target problem
introduced in \cite{SoTo02a,SoTo02b}.
The model with transaction costs is introduced in Section \ref{sec: model with frictions} and the corresponding dynamic programming
equation is a quasi-variational inequality \reff{eq: dpe v eps}.

The main result of the paper, outlined
in Section \ref{sec: small transaction cost expansion main result},
is the asymptotic expansion \reff{eq: expansion}.
It is proved under the hypothesis of Theorem \ref{thm: main} and
 states that the loss due to frictions
is proportional to the 2/3 power of the
proportional transaction cost
and the coefficient of the first term
in the expansion is characterized. Although our result is proved for a single risk criteria, it can be generalized to the multi-criteria case by exactly 
following the steps of \cite{BoVu12}. 
This extension naturally
increases the dimension of the corresponding PDE 
but does not introduce any additional technical difficulties.

In the case of  exponential and power utility functions, $\Psi$,
explicit formulae are available.  We collect them in Section \ref{sec: example}. In Section \ref{sec: proof examples}, we also explain how to construct almost optimal strategies.

In particular, if one chooses the threshold $p$ to be
the value function of the same utility
maximization problem with transaction costs
but without any liability,
one recovers the utility
indifference price  and its asymptotics.
In this context this price was first studied by
  \cite{DaPaZa93}.
In the case of
an exponential utility,
they obtained the price
as the difference of two functions.
These functions are related
to the maximum utility of two similar
problems whose solutions are described through
a nonlinear parabolic equation
with gradient constraints.
Related asymptotic formulae were formally
derived in \cite{WhWi97} and only recently
were proved rigorously by Bichuch in \cite{Bichuch11}.
Later
\cite{PoRo13} used an approach similar
to ours for this problem.
As discussed above,
the problem we study is equivalent
to hedging the option not perfectly
but with a prescribed expected loss.
As a consequence, our results
described in Section \ref{sec: example}
yield the asymptotic formula of  \cite{Bichuch11}.

The paper is organized as follows.
The next section describes the model
and its frictionless counterpart.
In Section \ref{sec: small transaction cost expansion main result},
we state the main theorem and our assumptions.
We illustrate this result in the cases of exponential
and power utilities in Section \ref{sec: example}.
Section \ref{sec: proof main thm} is devoted to the
proof of the main theorem and Section \ref{sec: proof examples}
verifies the assumptions in the examples.
In Section
\ref{sec: proof lem existence and property varpi},
we prove several technical estimates.

{\bf Notations:} Given $\Oc\subset \R^{k}$ and a smooth function $\vp:(t,x^{1},...,x^{k})\in [0,T]\x \Oc\mapsto \R$,  we write $\vp_{t}$ and $\vp_{x^{i}}$ for the partial derivatives with respect to $t$ and $x^{i}$. Second order derivatives are denoted by $\vp_{x^{i}x^{j}}$, and so on... We use the notations $D\vp$ and $D^{2}\vp$ to denote the gradient and the Hessian matrix with respect to the space component $(x^{1},...,x^{k})$. If we want to define them with respect to a subfamily, say $(x^{1},\cdots,x^{i})$, we write $D_{(x^{1},\cdots,x^{i})}\vp$ and $D^{2}_{(x^{1},\cdots,x^{i})}\vp$. When $\vp$ depends on only one variable, we simply write $\vp'$ and $\vp''$ for the first and second order derivatives. Any element of $\R^{k}$ is viewed as a column vector, and $^{\top}$ denotes the transposition. For an element $\zeta\in\R^k$ and $r>0$,
the open ball of radius $r>0$ centered at $\zeta$ is denoted by $B_{r}(\zeta)$. We let $\bar B$ and ${\rm Int}(B)$ denote the closure and the interior of $B$. Assertions involving random variables have to be understood in the a.s.~sense, if nothing else is specified.

\section{Partial hedging under expected loss constraints and pricing equations}\label{sec: models}

As usual, we let $(\Omega, \Fc,\P)$ be a complete probability space supporting a one dimensional Brownian motion $W$,  $\F:=(\Fc_{t})_{t\le T}$ be 
the right-continuous augmented filtration generated by $W$
and $T>0$ be the fixed time horizon.

\subsection{Controlled loss pricing with  proportional transaction costs}\label{sec: model with frictions}

We consider a financial market which consists of a single risky asset $S$, called stock hereafter. For ease  of notations, we assume that the risk free interest rate is $0$.
 Given initial data $(t,s)\in [0,T]\x (0,\infty)$, we let $S^{t,s}$ describe the evolution of this asset, and we assume that it follows the dynamics
\be\label{eq: def S}
S^{t,s}=s+\int_{t}^{\cdot} S^{t,s}_{\tau}\mu(\tau,S^{t,s}_{\tau})d\tau +\int_{t}^{\cdot}S^{t,s}_{\tau}\sigma(\tau,S^{t,s}_{\tau})dW_{\tau},
\ee
in which
\be\label{eq: hyp mu sigma}
&(t,s)\in [0,T]\x (0,\infty) \mapsto (s\mu(t,s),s\sigma(t,s))\in \R\x (0,\infty)&\\
& \mbox{ is Lipschitz continuous in $s$ and continuous in $t$.}&\nonumber
\ee
The latter condition implies the  existence and uniqueness of a strong solution.

\vs2

Transactions on this market are subject to a proportional   cost\footnote{See \cite{KaSa09} for a general presentation of models with proportional transaction costs.} described by a parameter $\eps^{3}>0$. We use the notation $\eps$ because we will be interested by the asymptotic $\eps\to 0$. The scaling $\eps^{3}$ is just for notational convenience, as it will be clear later on.

As usual in the presence of transaction costs, a portfolio process has to be described by a two dimensional process $(Y,X)$ in which $Y$ denotes the cash account and
$X$ denotes the amount of money invested in the stock.
We therefore call $(y,x)\in \R^{2}$ an initial endowment  at time $t$ if  $y$ is the position in cash  and $x$ is the amount    invested in the stock  at time $t$. Then, a financial strategy is an    adapted process  $L$ with bounded variations. The quantity $L_{\tau}-L_{t-}$ must be interpreted as the cumulated amount of money transferred on the time interval $[t,\tau]$ from the cash account    into the account invested in the stock.  It admits the canonical decomposition into two non-decreasing adapted processes $L=L^{+}-L^{-}$.  We denote by $\Lmf$ the collection of trading strategies.

Given an initial endowment $(y,x)$ at time $t$,   the portfolio process $(Y^{t,y,\eps,L},$ $X^{t,x,s,L})$ associated to the strategy $L\in \Lmf$ evolves according to
   \b*
        &\displaystyle
            Y^{t,y,\eps,L} = y- \int_t^\cdot (1+\eps^{3})dL^{+}_{\tau} + \int_t^\cdot (1-\eps^{3})dL^{-}_{\tau}\;,
        &\\
        &\displaystyle
            X^{t,x,s,L} = x+ \int_t^\cdot X^{t,x,s,L}_{\tau}\frac{dS^{t,s}_{\tau}}{S^{t,s}_{\tau}} + \int_t^\cdot dL^{+}_{\tau}-\int_t^\cdot dL^{-}_{\tau}.
   \e*
In order to rule out any possible arbitrage, we restrict the set of admissible strategies to the elements    of $\Lmf$  such that  the liquidation value of the portfolio is bounded from below, i.e. $L\in \Lc$ is admissible if there exists $c_{L}\ge 0$ such that
\be\label{eq: admissibility condition with transaction costs}
 Y^{t,y,\eps,L} +\ell^{\eps}(X^{t,x,s,L}) \ge -c_{L} \mbox{ on $[t,T]$},
\ee
where
$$
\ell^{\eps}:r\in \R\mapsto r-\eps^{3}|r|.
$$
 We denote by $ \Lmf^{\eps}(t,s,y,x)$ the set of admissible strategies associated to the initial data $(s,y,x)$ at time $t$.

\vs2

We now consider a trader whose  aim  is to hedge a plain vanilla European option with payoff function $g:r\in (0,\infty) \mapsto g(r)\in \R$. Hereafter, $g$ is assumed to be continuous with linear growth. In general, super-hedging in the presence of proportional transaction costs is much too expensive to make sense in practice, see    \cite{CvPhTo99,LeventalSkorohod97,SonerShreveCvitanic95}, and  \cite{BoTo00} for the multivariate setting.
We therefore introduce a {\sl risk criteria} under which the pricing and the hedging of the option will be performed.
It is specified through a map  $\Psi: r\in \R \mapsto \Psi(r)\in (-\infty,0]$, which we call {\sl loss function}.  We assume that $\Psi$ is concave{\footnote{We make this assumption
to obtain the representation in Proposition \ref{prop: explicit formula pi}.
This representation is then used
to verify the assumptions.  Hence,
the main result applies to general
loss functions provided the assumptions are verified.}}, non-decreasing, continuous on its domain, that   ${\rm Im}(\Psi):=\{\Psi(r),\;r\in \R$ s.t. $\Psi(r)>-\infty\}$ is open and  that
$$
\Esp{\Psi(-g(S^{t,s}_{T}))}>-\infty \;\;\;\mbox{ for all } (t,s)\in [0,T]\x (0,\infty).
$$

The hedging price associated to the loss function $\Psi$ and a {\sl threshold} $p\in  {\rm Im}(\Psi)$ is then defined by
\be\label{eq: def veps}
v^{\eps}(t,s,p,x):=\inf\left\{y\in \R:\exists\; L\in \Lmf^{\eps}(t,s,y,x)\mbox{ s.t. } \Esp{\Psi\left(
\Delta^{\eps,L}_{t,s,y,x}\right)}\ge p\right\},
\ee
where
$$
\Delta^{\eps,L}_{t,s,y,x}:=Y^{t,y,\eps,L}_{T}+\ell^{\eps}(X^{t,x,s,L}_{T})-g(S^{t,s}_{T}).
$$
The value $v^{\eps}(t,s,p,x)$ is the minimal initial price at which the option with payoff $g(S^{t,s}_{T})$ should be sold in order to ensure that the expected loss, as evaluated through $\Psi$, is not below the threshold $p$. Note that the assumption that $\Psi$ is bounded from above is rather natural since we consider here a risk criterion,
i.e.~one should not have   the possibility of compensating losses by unbounded gains. From the mathematical point, it could be relaxed up to additional integrability conditions ensuring that the corresponding optimization problem Max~$\E[{\Psi(
\Delta^{\eps,L}_{t,s,y,x})}]$ over $L\in \Lmf^{\eps}(t,s,y,x)$ is well-posed, see e.g. \cite{Bo02b} and the references therein.
Also note that this problem is of interest even in the {\sl degenerate} case $g\equiv 0$. Then, $v^{\eps}$ represents the threshold under which the cash account should not go in order for the terminal wealth to satisfy the requirement in \reff{eq: def veps}. This threshold is a building block for the analysis of optimal investment problems under risk constraints, see \cite{BoElIb10,bouchard2012weak}.
\\

The problem \reff{eq: def veps} is  a {\sl stochastic target problem with controlled loss} in the terminology of \cite{bet09}.  In order to obtain a pde characterization, the first step of their analysis consists of increasing the dimension of the state space and of the set of controls in order to turn the target problem under controlled loss in \reff{eq: def veps} into a target problem with $\P$-a.s. terminal constraint in the form of \cite{SoTo02a,SoTo02b}.  Namely,
$v^{\eps}$ admits the equivalent formulation
\be\label{eq: def alternative veps}
v^{\eps}(t,s,p,x)=\inf\left\{y\in \R:\exists\; (L,\alpha)\in \Lmf^{\eps}(t,s,y,x)\x\Amf \mbox{ s.t. }  \Psi\left(
\Delta^{\eps,L}_{t,s,y,x}\right)\ge P^{t,p,\alpha}_{T}\right\}\;,
\ee
where $\Amf$ denotes the set of a.s.~square integrable predictable processes such that
\be\label{eq: def P}
P^{t,p,\alpha}:=p+\int_{t}^{\cdot} \alpha_{\tau} dW_{\tau}\;\;\;\mbox{ is a martingale on $[t,T]$.}
\ee
One direction follows by taking expectation, the other one is just a consequence of the martingale representation theorem applied to $ \Psi(
\Delta^{\eps,L}_{t,s,y,x})$. Since  ${\rm Im}(\Psi)$ is convex, by the continuity of $\Psi$ on its domain, it is not difficult to see that we can even restrict the martingale $P^{t,p,\alpha}$ to take values in  ${\rm Im}(\Psi)$, see \cite{bet09,moreau10}.

Note that this reformulation is natural. Indeed, the expectation in \reff{eq: def veps} has to be understood as a conditional expectation given the (trivial) information at the starting point $t$. The conditional expectation evolves as time passes, and has no reason to stay above the initial threshold $p$. The martingale process $P^{t,p,\alpha}$ is here to take this evolution into account and turns the problem into a time-consistent one: it describes the evolution of the conditional expectation of  $\Psi(
\Delta^{\eps,L}_{t,s,y,x})$.

A {\sl geometric dynamic programming principle} for  problems of the form \reff{eq: def alternative veps}  was first obtained by \cite{SoTo02a,SoTo02b}. In the present framework, in which controls are of bounded variation, it was further studied by \cite{BoDa10}. Up to slight modifications, see the Appendix, it follows from the analysis in  \cite{BoDa10} that $v^{\eps}$ is a (discontinuous) viscosity solution on
    $
    \D\x\R
    $
of
\beq\label{eq: dpe v eps}
       \begin{array}{c} \max
        \left\{
            - \LSX \vp -   \LPlSXh \vp\;,\;
            -\eps^{3}+1+ \vp_{x}\;,\;
            -\eps^{3}-(1+ \vp_{x})
        \right\}=0\;\mbox{ on } \Di\x \R\;, \\
   \Psi( \vp+x-\eps^{3}|x|-g)=p \;\mbox{ on } \DT\x \R\;,
    \end{array}
    \eeq
in which we use the notations
    $$
     \Di:=[0,T)\x(0,\infty)\x {\rm Im}(\Psi)\;\;\mbox{ , }\;\DT:=\{T\}\x (0,\infty)\x {\rm Im}(\Psi)\;\;\mbox{ , }\;\;
     \D:=\Di\cup \DT,
    $$
    and
        \b*
        &\displaystyle \LPlSX^{a}\vp := \textstyle{ \frac12 }  a (\bar \sigma_{a} +\bar \sigma_{0})^{\top} D\vp_{p}\;,&\\
          &\displaystyle   \LPlSXh\vp :=\inf\{\LPlSX^{a}\vp :  a\in\R \st \bar \sigma_{a}^{\top} D \vp =0 \}\;,&\\
        &\displaystyle \LSX\vp :=
             \vp_{t}+\bar \mu^{\top} D \vp +\textstyle{ \frac12 }\Tr{\bar \sigma_{0}\bar \sigma_{0}^{\top} D^{2}\vp }\;,&
    \e*
  where  $D\vp_{p}$ is vector of the derivatives of the partial derivative $\vp_{p}$
  and for a given point $(t,s,x,a)\in [0,T]\x (0,\infty)\x \R\x \R$,
  \be\label{eq: def bar sigma a}
  \bar \mu(t,s,x):=\left(\begin{array}{c}s\mu(t,s)\\ x\mu(t,s)\\0\end{array}\right)
  \;\;\mbox{ and } \;\; \bar \sigma_{a}(t,s,x):=\left(\begin{array}{c}s\sigma(t,s)\\ x\sigma(t,s)\\a\end{array}\right)\;.
  \ee

 \begin{theorem}\label{thm: pde veps} Assume that $v^{\eps}$ is locally bounded. Then, it is a discontinuous viscosity solution of \reff{eq: dpe v eps}.
    \end{theorem}

 The above characterization can be exploited to compute the pricing function $v^{\eps}$ numerically. However, it should be observed that the operator $\LPlSXh$ involves an optimization over the unbounded set $\R$, which makes it discontinuous, and possibly difficult to handle numerically. Moreover, except if $v^{\eps}$ is smooth, the above pde does not allow to recover the associated hedging strategy.
 \vs2

 In this paper, we follow the approach of  \cite{SoTo12}, and try to provide an expansion of $v^{\eps}$ around $\eps=0$, i.e. for small values of the transaction costs.  For $\eps=0$, the financial market is complete and the problem can be solved explicitly by tools from convex analysis as described in the next subsection.
 We can therefore hope to obtain an explicit expansion, or at least a characterization of the different terms in the expansion  which will be more tractable from the numerical point of view.

\subsection{The frictionless benchmark case}\label{sec: frictionless model}

	We now consider the frictionless case  which   will be used to provide an expansion of $v^{\eps}$. We refer to \cite{FoLe99,FoLe00} for a general exposition of quantile and loss hedging problems in this context, see also \cite{BER13}.
	\vs2
	
	Let $\Uc$ denote the set of $\R$-valued progressively-measurable and a.s.~square integrable processes. Elements of $\Uc$ will be interpreted as amounts of money invested in the risky asset $S$. Given an initial allocation in amount of cash $z$ at time $t$ and $\vc\in \Uc$, the corresponding (frictionless) wealth process  $Z^{t,s,z,\vc}$ evolves according to
        $$
            Z^{t,s,z,\vc}  = z + \int_t^\cdot \vc_\tau dS^{t,s}_\tau/S^{t,s}_{\tau},
        $$
and the analog of $v^{\eps}(t,s,p,0)$ in \reff{eq: def veps} is
        $$
            \pi(t,s,p) := \inf
                \left\{
                    z\in\R : \Esp{\Psi\left( Z^{t,s,z,\vc}_{T} - g(S^{t,s}_{T}) \right)}\ge p
                    \mbox{ for some } \vc\in \Uc(t,s,z)
                \right\},
        $$
in which $\Uc(t,s,z)$ is the restriction to controls $\vc \in \Uc$ such that
$$
Z^{t,s,z,\vc} \ge -c_{\vc} \;\mbox{ on $[t,T]$ for some $c_{\vc}\ge 0$}.
$$

Because this frictionless financial market is complete, one can describe $\pi$ explicitly under mild regularity and integrability conditions. We provide the proof of the following in the Appendix for completeness.

\begin{proposition}\label{prop: explicit formula pi} Fix $(t,s,p)\in \D$. Assume that the function $\Psi:\R\mapsto {\rm Im}(\Psi)$ is   invertible, and that its  inverse $\Phi$ is $C^{1}({\rm Im}(\Psi))$.   Assume further that $\Phi': {\rm Im}(\Psi) \to (0,\infty)$  admits an inverse $I$. Finally assume that $\lambda^{t,s}:=(\mu/\sigma)(S^{t,s})$ is square integrable and that the process $Q^{t,s}$ defined by
$$
        Q^{t,s}:=\exp\left\{\frac12\int_{t}^{\cdot} |\lambda^{t,s}_{\tau}|^{2} d\tau + \int_{t}^{\cdot} \lambda^{t,s}_{\tau} dW_{\tau}\right\}
$$
satisfies
$$
\E\left[I(\hat q  Q^{t,s}_{T})\right]=p\;\;\mbox{ for some $\hat q>0$,}
$$
and
$$
 g(S^{t,s}_{T}) + \Phi\circ I(\hat q Q^{t,s}_{T})  \in L^{1}(\Q^{t,s}) \;\mbox{ where }\; d\Q^{t,s}/d\P=1/Q^{t,s}_{T}.
$$
Then,
\be\label{eq: explicit formula pi in general case}
\pi(t,s,p)=\E^{\Q^{t,s}}\left[ g(S^{t,s}_{T}) + \Phi\circ I(\hat q Q^{t,s}_{T})\right].
\ee
\end{proposition}

As for the case with frictions, one can also  obtain a characterization of $\pi$ in terms of a suitable Hamilton-Jacobi-Bellmann equation, see \cite{bet09} and the Appendix. As in \cite{SoTo12}, it will be used to obtain an expansion of $v^{\eps}$ around $\eps=0$. We state it in terms of the function
       \be\label{eq: def v in terms of pi}
        v:(t,s,p,x)\in \D\x \R\mapsto\pi(t,s,p)-x,
        \ee
which is the analog of $v^{\eps}$ when the initial amount $x$ invested in the stock is non-zero.
We note that formally $v^{0}$, obtained by setting $\epsilon$ to zero, is equal to $v$.
In the following, we restrict to the case where $v$ is smooth, increasing and  strictly convex in the $p$ parameter (the monotony and convexity just follow from the monotony and  concavity of $\Psi$). A similar result in the sense of viscosity solutions can be found in \cite{bet09}.

 \begin{theorem}\label{thm: pde v} Assume that $\pi\in C^{1,2}(\Di)$ and that $\min\{\pi_{p},\pi_{pp}\}>0$ on $\Di$. Then,
 $v(t,x,p,x)=\pi(t,s,p)-x$ is a strong solution of
          \beq\label{eq: PDE pi}
            -\LStheta v -\LPlSthetah v =0\;\mbox{ on }\; \Di\x \R\;\mbox{ and }\; \Psi(v+x-g)=p\mbox{ on } \DT\x \R,
        \eeq
 where
          \b*
        &\displaystyle \LPlStheta^{a}\vp := 2^{-1}a (   \bar \sigma_{\theta,a} +\bar \sigma_{\theta,0})^{\top} D \vp_{p}\;,&\\
          &\displaystyle   \LPlSthetah\vp :=\inf\{\LPlStheta^{a}\vp :  a\in\R \st \bar \sigma_{\theta,a}^{\top} D \vp =0 \}\;,&\\
        &\displaystyle \LStheta\vp :=
            \vp_{t}+\bar \mu_{\theta}^{\top} D \vp + \frac12 \Tr{\bar \sigma_{\theta,0}\bar \sigma_{\theta,0}^{\top} D^{2}\vp }\;,&
    \e*
  with, for $(t,s,p)\in \D$,
  \be\label{eq: def bar sigma theta a}
  \bar \mu_{\theta}(t,s):=\left(\begin{array}{c}s\mu(t,s)\\ \theta(t,s,p)\mu(t,s)\\0\end{array}\right)
\;  \mbox{ and }\;  \bar \sigma_{\theta,a}(t,s,p):=\left(\begin{array}{c}s\sigma(t,s)\\ \theta(t,s,p)\sigma(t,s)\\a\end{array}\right)\;,
  \ee
        \beq\label{eq: def theta}\bal
            \theta(t,s,p)
            &= \left(s\pi_s +\frac{\pi_p}{\sigma}\frac{\left( \frac\mu\sigma \pi_p-\sigma s\pi_{sp} \right)}{\pi_{pp}}\right)(t,s,p)\;.
        \eal\eeq
\end{theorem}

\begin{remark} For later use, note that
        \be\label{eq: def hat a}
         \LPlSthetah v=\LPlStheta^{\hat a} v \mbox{ with } \hat a:= -\bar \sigma_{\theta,0}^{\top} D v/v_{p} =\frac{\frac{\mu}{\sigma}v_{p}- \sigma sv_{ps}}{v_{pp}}\;,
        \ee
        and
                \beq\label{eq: relation thetat with hat a}\bal
            \theta
            &=s\pi_{s}+\pi_{p}\hat a/\sigma\;.
        \eal\eeq
        \end{remark}

\section{Small transaction costs expansion}\label{sec: small transaction cost expansion main result}

It follows from Proposition \ref{prop: explicit formula pi} that the value function $v$ associated to the frictionless case is known, or at least can be computed easily. Since it should identify to $v^{\eps}$ for $\eps=0$, we seek for an expansion of $v^{\eps}$ as $\eps\to 0$ in which $v$ is the $0$-order term. From \cite{SoTo12}, one can expect to obtain an $o(\eps^{2})$-expansion if we introduce a second and a fourth order term, the last one depending on a {\sl fast variable}, $\xib_{\eps}$ below. Namely, we seek for two functions $u$ and $\varpi$ such that
    \be\label{eq: expansion}
    v^{\eps}(\zeta,x)=v(\zeta,x)+\eps^{2}u(\zeta)+\eps^{4}\varpi\circ\xib_{\eps}(\zeta,x)+o(\eps^{2})\;\mbox{ for } (\zeta,x)\in \D\x \R,
    \ee
    in which, for  a map $w:(\zeta,\xi)\in \D \x \R \mapsto w(\zeta,\xi)$, we set
    \be
    (w\circ\xib_{\eps})(\zeta,x):= w(\zeta,\xib_{\eps}(\zeta,x)),
    &\mbox{
    with}&
     \label{eq: def xi}
    \xib_{\eps}(\zeta,x):=\frac{x-\theta(\zeta)}{\eps}.
    \ee
Note that when $w$ has sub quadratic growth in $\xi$, the term $\eps^{4}\varpi\circ\xib_{\eps}(\zeta,x)$ in \reff{eq: expansion}
is in a lower order than $\eps^{2}$ and plays no role in the expansion.  We will show that this is indeed the case.  However, at least
at the formal level, the second derivative of $\eps^{4}\varpi\circ\xib_{\eps}(\zeta,x)$ is exactly of order $\epsilon^2$ and this observation is crucial in deriving the corrector equations.  Also, 
in the context of formal matched asymptotics,
one may recognize \reff{eq: expansion} as in the inner expansion.

    \begin{remark}\label{rem: bounded domain and claims} In the case where the domain of $\Psi$ is bounded from below, the convergence $v^{\eps}\to v$ can not hold except if $g$ is linear. Indeed, assume that  the domain of $\Psi$ is bounded by $-\kappa\in \R$, i.e. $\Psi\equiv -\infty$ on $(-\infty,-\kappa)$.  Then,  it follows from \cite{BoTo00} that $v^{\eps}(t,s,p,x)\ge \hat g(s)-x-\kappa$ for all $(t,s,p,x)\in \Di\x\R$, where $\hat g$ is the concave envelope of $g$. On the other hand $\lim_{t\to T}v(t,s,p,x)=g(s)-x-\kappa+\Psi^{-1}(p)+\kappa$, by \reff{eq: PDE pi},   where $\Psi^{-1}$ is the left-continuous inverse of $\Psi$. If  $g$ is not concave, i.e.~if $\{\hat g>g\}$ is non-empty,  we therefore  obtain that  $v^{\eps}$ does not converges to $v$ on a non-empty subset of $\{(t,s,p)\in \Di:\hat g(s)>g(s)+\Psi^{-1}(p)+\kappa\}$.  Hence, we need to assume that $g$ is concave, i.e. $\hat g\equiv g$.
It can actually neither be strictly concave on any interval of $(0,\infty)$. Otherwise, there will be $(t,s)$ such that $\E^{\Q^{t,s}}\left[ g(S^{t,s}_{T})\right]=:\bar \pi(t,s)<g(s)$ and therefore    $v(t,s,p,x)<g(s)-x+\Psi^{-1}(p)$ $=$ $g(s)-x-\kappa+\Psi^{-1}(p)+\kappa$, since adding $-x+\Psi^{-1}(p)$ to $\bar \pi(t,s)$ allows to hedge $Z_{T}:=g(S^{t,s}_{T})+\Psi^{-1}(p)$ which satisfies $\Psi(Z_{T}-g(S^{t,s}_{T}))=p$. By choosing $p$ such that $\Psi^{-1}(p)+\kappa$ is close to $0$, we again obtain that $v^{\eps}(t,s,p,x)$ does not converge to $v(t,s,p,x)$ even if $\hat g=g$
{\footnote{When the lower bound is zero,
the boundary of the natural domain of the problem is given
by the super-replication cost.  We believe that in this
case there is a boundary layer near this boundary.}}.
    \end{remark}

Our main result provides a precise characterization of the functions $u$ and $\varpi$ under the assumption that $v^{\eps}$ converges at a rate $O(\eps^{2})$.  We shall see that this is true  in typical  examples of application in Section  \ref{sec: example} below{\footnote{This assumption states
 that the expansion in the small parameter $\eps$
 starts with a quadratic term.  In other words,
 we assume that the order of proposed expansion is ``correct''.
 Under this and other regularity
 assumptions, we prove the expansion and
 derive formulae for the coefficients
 in the expansion.
 Indeed this assumption holds in many examples.
 However, in the case discussed in the Remark
 \ref{rem: bounded domain and claims}
 we believe that there is a boundary layer and this
 assumption would only hold
 away from the super-replication cost.}}.

 \begin{assumption}
 \label{ass: u eps finite}     For any $(\zeta_{o},x_o)\in\D\x\R$,
      there exists $r_o, \eps_o>0$
      such that
      \beq\label{eq: definition upper bound}
   \sup \left\{
            u^\eps(\zeta,x):=\frac{v^\eps(\zeta,x)-v(\zeta,x)}{\eps^2},\quad (\zeta,x)\in  B_{r_o}(\zeta_o,x_o)\cap (\D\x \R), \eps\in(0,\eps_o]
        \right\}<\infty.
      \eeq
     \end{assumption}

It allows us  to give a sense to the relaxed semi-limits
    \be\label{eq: def u star}
        u^*(\zeta,x) := \Limsup_{\eps\downarrow0,(\zeta',x')\rightarrow(\zeta,x)}  u^\eps(\zeta',x')
      \;  \mbox{ and }\;
        u_*(\zeta,x) := \Liminf_{\eps\downarrow0,(\zeta',x')\rightarrow(\zeta,x)}  u^\eps(\zeta',x'),
    \ee
which will be the main objects of our analysis. More precisely, we shall show that $u^{*}=u_{*}=:u$ does not depend on the $x$-variable and is a viscosity solution of
  \beq\label{eq: 2nde corrector equation}\left\{\begin{matrix}
        -\Hc \vp - h =0 & \mbox{on }\Di\;,\lpt7
        \vp = 0 & \mbox{on } \DT,
    \end{matrix}\right.\eeq
    where
	\be\label{eq: def Hc}		
      \Hc \vp&=& \vp_t + \frac12 \sigma^2s^2 \vp_{ss}
                        +\frac12\left(\hat a\right)^2\vp_{pp} +\sigma s \hat a\vp_{sp}-\frac\mu\sigma \hat a \vp_p,
       \ee
in which  $\hat a$ is defined in \reff{eq: def hat a}, and $(\varpi,h)$ are the solution of the so-called  {\sl first corrector equation}, i.e. for each $(\zeta,\xi)\in \Di\x \R$:
\beq\label{eq: first corrector equation}
        \max \{
            -\frac12[\frac{\pi_{pp}}{\left(\pi_p\right)^2}\sigma^2](\zeta)\xi^2
            +h(\zeta) - \frac12[\sigma^2\delta^2](\zeta)\varpi_{\xi\xi}(\zeta,\xi);
            -1+\varpi_\xi(\zeta,\xi);
            -1-\varpi_\xi(\zeta,\xi)
         \}=0,
    \eeq
    where
    \be\label{eq: def delta}
         \delta:=s\theta_s-\theta+\frac{\theta_p}{\pi_p}\left( \theta-s\pi_s \right).
      \ee

In order to construct the pair $(\varpi,h)$, we need some smoothness and non-degeneracy conditions on the value function $\pi$ of the frictionless problem.

     \begin{assumption}\label{ass: pi smooth} The functions $\pi,\theta$ and $\delta$ are $C^{1,2}(\D)$ and
      $        (\pi_{pp}\wedge\pi_p\wedge |\delta| ) >0$ on  $\D$.
    \end{assumption}

      \begin{lemma}\label{lem: existence and property varpi} Let the Assumption \ref{ass: pi smooth} hold. Then, there exists a locally bounded function $h$ on $\D$   and  a non-negative function $\varpi$ on $\D\x \R$ such that, for all  $\zeta\in \D$, the map $\xi \in \R\mapsto \varpi(\zeta,\xi)$ is $C^{2}(\R)$   and  solves \reff{eq: first corrector equation} on $\R$.
          Moreover, it satisfies
      \benumlabi{}
      \item $\varpi(\cdot,0)=0$   on $\D$.
      \item $\varpi\in C^{1,2}(\D\x \R)$ and
      $|\varpi_{\xi}|\le 1$ on $\D\x \R$.
      \item There exists a continuous function $\varrho:\D\to \R$ such that
      \be\label{eq: estimate varpi}
      \frac{|\varpi(\cdot,\xi)|}{1+|\xi|}+(|\varpi_{t}|+|D\varpi|+|D^{2}\varpi|)(\cdot, \xi)\le \varrho \mbox{ on } \D,\;\forall\;\xi \in \R.
      \ee
      \item There exists a continuous  positive function $ \hat \xib$ on $\D$ such that, for all $(\zeta,\xi)\in \D\x \R$,
      $$
      \varpi_{\xi}(\zeta,\xi)=-1 \;\Leftrightarrow\;\xi\le -\hat  \xib(\zeta)\; \mbox{ and }\;  \varpi_{\xi}(\zeta,\xi)=1 \;\Leftrightarrow\;\xi\ge  \hat  \xib(\zeta).
      $$
      \end{enumerate}
      \end{lemma}
The proof of this result is postponed to Section \ref{sec: proof lem existence and property varpi}.
In that section, we also derive  explicit expressions for $\varpi$, $h$ and $\hat \xib$ in terms of $\pi$ and its derivatives, see      \reff{eq: explicit form of varpi}, \reff{eq: explicit form of h} and \reff{eq: explicit form of xib} below.

\begin{remark}\label{rem: property varpi} It follows from  Lemma \ref{lem: existence and property varpi} that we indeed have $|\varpi(\cdot,\xi)|\le |\xi|$ for all $\xi \in \R$. This is a straightforward consequence of  \reff{eq: first corrector equation} and (i).
\end{remark}

In order to fully characterize  $u$ as $u^{*}=u_{*}$, we also need a comparison principle on \reff{eq: 2nde corrector equation}.
    \begin{assumption}\label{ass: comparison for u} There exists a set of functions $\Cc$  which contains $u^{*}$ and $u_{*}$, and such that $u_1\ge  u_2$ on $\D$ whenever $u_{1}$ (resp. $u_{2}$) is a lower semi-continuous (resp. upper semi-continuous) viscosity super-solution (resp. sub-solution) of \reff{eq: 2nde corrector equation} in $\Cc$.
    \end{assumption}

Under the above conditions, we will prove in Section \ref{sec: proof main thm} that the expansion announced in   \reff{eq: expansion} holds.

\begin{theorem}\label{thm: main} Let the Assumptions \ref{ass: u eps finite},  \ref{ass: pi smooth}  and \ref{ass: comparison for u} hold. Then,  \reff{eq: expansion} holds with $\varpi$ as in Lemma \ref{lem: existence and property varpi} and $u$ given by the unique viscosity solution of \reff{eq: 2nde corrector equation} in $\Cc$. Moreover, $u=u^{*}=u_{*}$.
\end{theorem}
\proof This is an immediate consequence of Propositions \ref{prop : subsol u*}, \ref{prop : supersol u*} and \ref{prop: terminal condition for u} below, combined with Assumption \ref{ass: comparison for u}.
\ep
\\

As explained above, the function $\pi$ is explicit  or can be computed easily, and so is $v$, while $\varpi$ is given in \reff{eq: explicit form of varpi} below in terms of $\pi$ and its derivatives. As for $u$,  it solves the linear equation \reff{eq: 2nde corrector equation} which can be solved numerically whenever the function  $\hat a$ defined in \reff{eq: def hat a} and $\hat a \mu/\sigma$ are Lipschitz on $\D$.  Note that, in this case,  it admits the Feynman-Kac representation
    $$
           u(t,s,p)=\E\left[\int_t^T h\left(\tau,\bar S^{t,s}_\tau,\bar  P^{t,s,p}_\tau \right)d\tau\right],
    $$
  in which $\bar S^{t,s}$ solves \reff{eq: def S} with $\mu\equiv0$,   and
  $$
  \bar  P^{t,s,p}:=p-\int_{t}^{\cdot} (\hat a\mu/\sigma)(\tau,\bar S^{t,s}_{\tau},\bar P^{t,s,p}_{\tau})d\tau+\int_{t}^{\cdot} \hat a(\tau,\bar S^{t,s}_{\tau},\bar P^{t,s,p}_{\tau})dW_{\tau}.
  $$
  If the probability measure $\Q^{t,s}$ of Proposition \ref{prop: explicit formula pi} is well defined, this is equivalent to
    $$
          u(t,s,p)=\E^{\Q^{t,s}}\left[\int_t^T h\left( \tau,  S^{t,s}_\tau,\hat  P^{t,s,p}_\tau \right)d\tau\right],
    $$
  in which
  $$
  \hat   P^{t,s,p}:=p+\int_{t}^{\cdot} \hat a(\tau,S^{t,s}_{\tau},\hat P^{t,s,p}_{\tau})dW_{\tau}.
  $$
  In the examples of Section \ref{sec: example}, all these quantities are known, as far as one can compute the price and the greeks of a plain vanilla European option in the Black and Scholes model.

  Note also that the functions $\pi$ and $\hat \xib$ can be used to construct almost optimal strategies in the original problem \reff{eq: dpe v eps}. This will be explained later on   in    Section \ref{sec: proof examples} for the exponential and the power risk criterias.

  \begin{remark} We restrict here to the case of a single stock mainly for ease of notations. The  arguments contained in Section \ref{sec: proof main thm} can essentially be reproduced in the multidimensional case. The main difficulties will come from  the construction of $\varpi$ in Lemma \ref{lem: existence and property varpi}, see \cite{PoSoTo12}, and from the existence of a solution to the Skorohod problem in the proofs of Section \ref{sec: proof examples}.
  \end{remark}

\section{Examples}\label{sec: example}

In this section, we discuss two typical examples of application in which Assumptions \ref{ass: u eps finite},  \ref{ass: pi smooth}  and \ref{ass: comparison for u} are satisfied, and therefore the expansion result of Theorem \ref{thm: main} can be applied.

\subsection{The exponential risk criterion in the Black and Scholes model}

    We first specialize the discussion to the case where the loss function $\Psi$ is of exponential form:
    \be\label{eq: Psi expo}
    \Psi(r):=-e^{-\eta r} \;,\;\;r\in \R\;,
    \ee
   {for some $\eta>0$}, and the   stock price $S^{t,s}$ follows the Black and Scholes dynamics
    \be\label{eq: dyna BS}
        S^{t,s}=s+\int_t^\cdot \lambda\sigma S^{t,s}_\tau d\tau+\int_t^\cdot \sigma S^{t,s}_\tau dW_\tau\;,
    \ee
    for some $(\lambda,\sigma)\in\R\x(0,\infty)$.

    In this case, the pricing function $\pi$ can be derived explicitly. This is an easy consequence of Proposition \ref{prop: explicit formula pi}. {We recall that $h$ and $\hat \xib$ are given in \reff{eq: explicit form of h} and \reff{eq: explicit form of xib} below.}

    \begin{proposition}\label{prop: pi explicit BS example} For all $(t,s,p)\in \D:=[0,T]\x (0,\infty)\x (-\infty,0)$,
    \be\label{eq: pi explicit in BS example}
    \pi(t,s,p)=\bar \pi(t,s)+\check \pi(t,p)\;,
    \ee
    where
    $$
    \check \pi(t,p):=-\frac{\lambda^{2}(T-t)}{2\eta}-\frac1\eta \ln(-p)
    $$
    and
    $$
    \bar \pi(t,s):=\E^{\Q }\left[g(S^{t,s}_{T})\right]\; \mbox{ with }\;d\Q /d\P:=e^{-\frac{\lambda^{2}}{2}T-\lambda W_{T} }.
    $$
    Moreover, if $\bar \pi\in C^{0,2}([0,T]\x (0,\infty))$, then
      \be
\left\{    \begin{array}{c}
  \theta(t,s) =s\bar \pi_{s}(t,s)+\frac{\lambda}{\sigma \eta}\;,\;
    \delta(t,s)  ={s^{2}}\bar \pi_{ss}(t,s)-\frac{\lambda}{\sigma \eta}
    \;,\;
    \hat a(p)=-\lambda p\;,\\
   h(t,s)  =\left(\frac{3}{16}\right)^{\frac23}\sigma^{2}\eta^{\frac13}|\delta(t,s)|^{\frac43}\;,\;
    \hat \xib(t,s)=\left(\frac{3}{2\eta}\right)^{\frac13}|\delta(t,s)|^{\frac23}.
    \end{array}
    \right. \label{eq: expo BS case - explicit functions}
    \ee

    \end{proposition}

  They are well-defined under the conditions of Assumption \ref{ass: BS expo case}  below.
Note in particular that
\be\label{eq: do not depend on x and p}
\theta, \delta, h \mbox{ and } \hat \xib \mbox{  only depend on  $(t,s)$.}
\ee

  Moreover, the second corrector  equation \reff{eq: 2nde corrector equation} can be written as
      \beq\label{eq: 2nde corrector equation BS case expo}
      \left\{\begin{matrix}
       -\vp_t - \frac12 \sigma^2s^2 \vp_{ss}
                        -\frac{\lambda^{2}}{2}p^2\vp_{pp} +\sigma\lambda s p \vp_{sp}+\lambda^{2}p \vp_p - h =0 & \mbox{on }\Di\;,\lpt7
        \vp = 0 & \mbox{on } \DT.
    \end{matrix}\right.\eeq
     If $h$ is bounded, which will be the case under Assumption \ref{ass: BS expo case} below, it follows from standard arguments that
     \be\label{eq: def bar u example}
   {\hat u}:(t,s)\in [0,T]\x (0,\infty)\mapsto \E^{\Q}\left[\int_{t}^{T} h(\tau,S^{t,s}_{\tau})d\tau\right],
    \ee
   is   the unique viscosity solution of \reff{eq: 2nde corrector equation BS case expo} in the class of functions having polynomial growth, see \cite{CrIsLi92}.

  We now impose conditions under which Assumptions \ref{ass: u eps finite},  \ref{ass: pi smooth}  and \ref{ass: comparison for u} of Theorem \ref{thm: main} hold true.
  In particular, they are similar to
  the assumptions used in \cite[Assumptions 3.1 and 3.2]{Bichuch11}.
  {\footnote{These assumptions can be verified directly
  using the frictionless equation and assumptions
  on $g$.}}

    \begin{assumption}\label{ass: BS expo case} The following holds:
    \begin{enumerate}
    \item[{\rm a.}] $\bar \pi \in C^{1,4}(\D)$.
   \item[{\rm b.}]    There exists $K>0$ such that
    $$
{|g|+  |s\bar \pi_{s}|+    |s^{2} \bar \pi_{ss}| + |\delta^{-1}|+ |\theta_{t}|+ |s^{2}\theta_{ss}|} \le K\;\;\mbox{    on $\D$.}
   $$
    \end{enumerate}
    \end{assumption}

    Note that these conditions imply in particular that    ${\hat u},\varpi  \in C^{1,2}(\D)$, see \reff{eq: expo BS case - explicit functions} and  \reff{eq: explicit form of varpi} below for the exact expression of $\varpi$.

     \begin{proposition}\label{prop: ass ok for expo BS}  Let $\Psi$  be as in \reff{eq: Psi expo} and $S$ as in \reff{eq: dyna BS}. Then, Assumption \ref{ass: BS expo case}  implies Assumptions \ref{ass: u eps finite},  \ref{ass: pi smooth}  and \ref{ass: comparison for u} of Theorem \ref{thm: main}.
     \end{proposition}

The proof of this proposition is postponed to Section \ref{sec: proof examples}.

\begin{remark}\label{cor: almost strategy expo}{\bf [$\eps^{2}$-optimal strategies]} In the course of the proof of Proposition \ref{prop: ass ok for expo BS}, we shall explain how to construct strategies which are optimal at the  order $O(\eps^2)$, or $o(\eps^{2})$ under an additional regularity assumption,  for the problem with transaction costs, and which only depends on the knowledge of $v$, $\hat u$, $\varpi$ and $\theta$.  See   Propositions \ref{prop: strat eps opti expo} and   \ref{prop: strat eps opti expo eps cube} below. \end{remark}

Note that, as a by-product, our expansion allows one to recover the result of \cite{Bichuch11} on the Hodges and Neuberger indifference price. More precisely, let $V^{\eps}$  be defined as
\b*
V^{\eps}(t,s,y,x)&:=&\sup_{ L\in \Lmf^{\eps}(t,s,y,x)}\Esp{\Psi\left( Y^{t,y,\eps,L}_{T}+\ell^{\eps}(X^{t,x,s,L}_{T})-g(S^{t,s}_{T})\right)}
\e*
and let $\tilde V^{\eps}$ be defined similarly but for $g\equiv 0$. Then, the indifference price associated to the market with   transaction costs is given by
\b*
q^{\eps}(t,s,y,x)&:=&\inf\{q\in \R:  V^{\eps}(t,s,y+q,x)\ge \tilde V^{\eps}(t,s,y,x)\}.
 \e*
 It is easy to see that, for the exponential risk criterion,  $q^{\eps}$ does not depend on the $y$-variable and that
 $$
 q^{\eps}(t,s,x)=-\frac{1}{\eta}\ln\left(\frac{\tilde V^{\eps}(t,s,y,x)}{V^{\eps}(t,s,y,x)}\right)=v^{\eps}(t,s,-1,x)-\tilde v^{\eps}(t,s,-1,x)\;,
 $$
 in which $\tilde v^{\eps}$ is defined as $v^{\eps}$ but for $g\equiv 0$. Under the assumptions of Proposition \ref{prop: ass ok for expo BS}, it then follows that
 $$
 q^{\eps}(t,s,x)=  \bar \pi(t,s) +\eps^{2}\E^{\Q}\left[\int_{t}^{T} \Delta h(\tau,S^{t,s}_{\tau})d\tau\right]+o(\eps^{2})\;,
 $$
 in which
 $$
 \Delta h(t,s):=\left(\frac{3}{16}\right)^{\frac23}\sigma^{2}\eta^{\frac13}\left(|\delta(t,s)|^{\frac43}-\left|\frac{\lambda}{\sigma \eta}\right|^{\frac43}\right).
 $$

\subsection{The power risk criterion in the Black and Scholes model}
We now consider the case
\be\label{eq: psi power}
\Psi(r):=-(r+\kappa)^{-\beta}\1_{\{r>\kappa\}}-\infty\1_{\{r\le \kappa\}}\;,\;\;r\in \R,
\ee
with $\beta,\kappa>0$.  For this risk function, Proposition \ref{prop: explicit formula pi} implies that $\pi=\bar \pi+\hat \pi$ with
\be\label{eq: explicit price power case}
\bar \pi(t,s)=\E^{\Q}[g(S^{t,s}_{T})]\;\mbox{ and }\; \hat \pi(t,p):=-\kappa+ (-p)^{-\frac{1}{\beta}}m(t)\;,
\ee
for $(t,s,p)\in \D$, in which $m$ is a $C^{1}_{b}([0,T])$ {positive} function satisfying $m(T)=1$.

In view of Remark \ref{rem: bounded domain and claims}, we can however not expect to have $v^{\eps}\to v$ if $g$ is not linear. Since any linear payoff is hedged perfectly by the same buy-and-hold strategy in the two models, this boils down to considering the case $g\equiv 0$ up to an initial shift of $\kappa$ and $x$,  at the costs of  an additional $\eps^{3}$ term. We therefore restrict to the degenerate case  $g\equiv 0$. Recall from Section \ref{sec: model with frictions} that the problem remains of interest, as $v^{\eps}$ is a building block for the analysis of optimal investment problems under risk constraints, see \cite{BoElIb10,bouchard2012weak}.

 \begin{proposition}\label{prop: ass ok for powe BS}   Let $\Psi$ be as in \reff{eq: psi power}, $S$ as in \reff{eq: dyna BS} and  $g\equiv 0$. Then Assumptions \ref{ass: u eps finite},  \ref{ass: pi smooth}  and \ref{ass: comparison for u} of Theorem \ref{thm: main} hold.
 \end{proposition}

The proof is postponed to Section \ref{sec: proof examples}.

\begin{remark}\label{cor: almost strategy power}{\bf [$\eps^{2}$-optimal strategies]} As in the exponential case, we produce in the course of the proof of Proposition \ref{prop: ass ok for powe BS} a strategy which is optimal at the   order $o(\eps^2)$ for the problem with transaction costs, and which only depends on the knowledge of $v$, $\hat u$, $\varpi$ and $\theta$.  See   Remark \ref{rem: strat eps opti power} below. \end{remark}

\section{Derivation of the small transaction costs expansion}\label{sec: proof main thm}

 \subsection{Preliminaries}	

 We start with the derivation of easy estimates that will be of important use in the sequel.

        \begin{remark}\label{rem: encadrement v eps}
       Observe  that, for $(\zeta,x)\in\D\x\R$, the initial dotation in cash and amount of stock $(v^{\eps}(\zeta,x)+x+\eps^{3}|x|,0)$ can be turned into  $(v^{\eps}(\zeta,x),x)$  by an immediate transfer $\Delta L_{0}=x$, while the initial dotation $(v^{\eps}(\zeta,0)-x+\eps^{3}|x|,x)$ can be turned into $(v^{\eps}(\zeta,0),0)$ by an immediate transfer $\Delta L_{0}=-x$. By the definition of $v^{\eps}$, this implies that
        \be\label{eq: encadrement v eps}
            v^\eps(\zeta,0)-\eps^3|x|
            \le v^\eps(\zeta,x)+x\le
            v^\eps(\zeta,0)+\eps^3|x|.
        \ee

        \end{remark}

\begin{remark}\label{rem: v le veps} It follows from the same arguments as in \cite[Proposition 6.1]{BoTo00} that
$
v^{\eps} \ge v .
$
\end{remark}

        \begin{lemma}\label{lem: u does not depend on x}
          \benumlabi{}
            \item The functions $u^*$ and $u_*$ are independent of the $x$-variable;
            \item Moreover, for all $\zeta\in \D$, we have
                \beq\label{eq: relaxed semi-limits theta}
                    u^*(\zeta) = \Limsup_{\eps\downarrow0,\zeta'\rightarrow\zeta}u^{\eps*}(\zeta',\theta(\zeta'))
                    \qquad\mbox{and}\qquad
                    u_*(\zeta) = \Liminf_{\eps\downarrow0,\zeta'\rightarrow\zeta}u^{\eps}_{*}(\zeta',\theta(\zeta'))\;,
                \eeq
                in which $u^{\eps*}$ and $u^{\eps}_{*}$ denote the upper- and lower-semicontinuous envelopes of $u^{\eps}$.
          \end{enumerate}
        \end{lemma}
        \proof We only show the result for $u^{*}$, the same reasoning can be used for the relaxed  semi-limit $u_{*}$.
        Fix $\zeta\in \D$ and  $x\in \R$.  By the definition of $u^*$, there exists a sequence $(\zeta_\eps,x_\eps)_{\eps>0}$ such that
            \beq\label{eq: convergence u eps u star u not x}
                (\zeta_\eps,x_\eps)\underset{\eps\downarrow0}{\longrightarrow}(\zeta,x)
                \quad\mbox{and}\quad
                u^\eps(\zeta_\eps,x_\eps) \underset{\eps\downarrow0}{\longrightarrow} u^*(\zeta,x).
            \eeq
            Fix also a sequence $(x'_\eps)_{\eps>0}$ going to $x'\in \R$ as $\eps\to 0$.
            By Remark \ref{rem: encadrement v eps} and the definitions of $  u^\eps$ and $v$ in \reff{eq: definition upper bound} and  \reff{eq: def v in terms of pi},
            we have
            \b*
                &\displaystyle v^\eps(\zeta_\eps,0)-\eps^3|x_\eps|
                \le \eps^2   u^\eps (\zeta_\eps,x_\eps) + \pi(\zeta_\eps) \le
                v^\eps(\zeta_\eps,0)+\eps^3|x_\eps|\;,&\\
                &\displaystyle v^\eps(\zeta_\eps,0)-\eps^3|x'_\eps|
                \le \eps^2   u^\eps (\zeta_\eps,x'_\eps) + \pi(\zeta_\eps) \le
                v^\eps(\zeta_\eps,0)+\eps^3|x'_\eps|\;,&
            \e*
            so that
            $$
                -\eps\left( |x_\eps|+|x'_\eps| \right)
                \le   u^\eps(\zeta_\eps,x_\eps)-  u^\eps(\zeta_\eps,x'_\eps) \le
                \eps\left( |x_\eps|+|x'_\eps| \right).
            $$
            Sending $\eps\rightarrow0$ and using \reff{eq: convergence u eps u star u not x} then leads to
            $$
                \lim_{\eps\rightarrow0}  u^\eps(\zeta_\eps,x'_\eps)=u^*(\zeta,x).
            $$
           This shows in particular  that  $ u^*(\zeta,x')\ge u^*(\zeta,x)$. By arbitrariness of $x,x'\in \R$, the reverse inequality holds as well, showing that $u^{*}$ does not depend on its $x$-variable. Moreover, applied to $x=x':=\theta(\zeta)$ and $x'_{\eps}:=\theta(\zeta_{\eps})$, the above implies that
           $$
            \Limsup_{\eps\downarrow0,\zeta'\rightarrow\zeta}u^{\eps*}(\zeta',\theta(\zeta'))\ge u^*(\zeta,\theta(\zeta))=
            \Limsup_{\eps\downarrow0,(\zeta',x')\rightarrow(\zeta,\theta(\zeta))}u^\eps(\zeta',x').
           $$
To conclude the proof of the left hand-side of \reff{eq: relaxed semi-limits theta}, it  remains to show that
           \be\label{eq: limsup through ueps* of theta}
            \Limsup_{\eps\downarrow0,(\zeta',x')\rightarrow(\zeta,\theta(\zeta))}u^\eps(\zeta',x')=\Limsup_{\eps\downarrow0,(\zeta',x')\rightarrow(\zeta,\theta(\zeta))}u^{\eps*}(\zeta',x'),
           \ee
           and to use the inequality
           $$
          \Limsup_{\eps\downarrow0,(\zeta',x')\rightarrow(\zeta,\theta(\zeta))} u^{\eps*}(\zeta',x')\ge \Limsup_{\eps\downarrow0,\zeta'\rightarrow\zeta}u^{\eps*}(\zeta',\theta(\zeta')).
           $$
     To see that the above holds, note that the continuity of $v$, see Assumption \ref{ass: pi smooth} and recall \reff{eq: def v in terms of pi}, implies that for $(\zeta,\xi) \in \D\x \R$ and $\eps>0$  we can
        find $(\zeta_{\eps},\xi_{\eps}) \in \D\x \R$ such that $ (v^{\eps}-v)(\zeta,\xi)\le   (v^{\eps*}-v)(\zeta,\xi)\le  (v^{\eps}-v)(\zeta_{\eps},\xi_{\eps})+\eps^{3}$. Recalling the definition of $u^{\eps}$ in \reff{eq: definition upper bound}, this proves \reff{eq: limsup through ueps* of theta}.

        \qed\\

        In view of the above result, we shall from now on omit the $x$-variable in the functions $u^{*}$ and $u_{*}$.

\subsection{The key expansion lemma}

    We now provide the following key  lemma, which is the counterpart of \cite[Remark 3.4, Section 4.2]{SoTo12}.

    \begin{lemma}\label{lem: remainder estimate} Assume that $\pi,\theta \in C^{1,2}(\Di)$.
    For  $\eps>0$, and two $C^{1,2}(\Di\x \R)$ functions $\phi$ and $w$, define
      \beq\label{eq: test function form for remainder estimate}
        \psi^\eps=v + \eps^2 \phi + \eps^4 w^{\eps}\; \mbox{ with }\;w^{\eps}:=w\circ \xib_{\eps}.
      \eeq
      Set $D^{\iota}_{\eps}:=(\Di\x \R)\cap \{\psi^{\eps}_{p}>0\}\cap \{\eps^2 \phi_{p} + \eps^4 w^{\eps}_{p}\ge \iota \pi_{p}\}$ for some $\iota>-1$.
      Then,
      \be\label{eq: lem expansion}
       \eps^{-2}  (\LSX + \LPlSXh)\psi^{\eps}
       =
      \frac12\frac{\pi_{pp}}{\left(\pi_p\right)^2}\sigma^2\xib_\eps^2
           + (\Hc+\LXlSP^{\hat a})\phi+ \frac12\sigma^2\delta^2(w_{\xi\xi}\circ\xib_{\eps}) + R_{\eps} \;\mbox{ on } D^{\iota}_{\eps},
      \ee
      where
        \b*
      \LXlSP^{\hat a}\phi &=&  \frac12 \sigma^2 \theta^2 \phi_{xx} + \sigma^2s\theta\phi_{sx} + \theta \sigma \hat a\phi_{px}
       \e*
with $\hat a$   defined in \reff{eq: def hat a}, and  where      $R_{\eps}$ is a continuous map defined on $D^{\iota}_{\eps}$ such that:
      \benumlabi{}
      \item   For each bounded set $B\subset D^{\iota}_{\eps}$, there exists $\eps_{B}>0$ such that $\{\eps^{-1}R_{\eps}(\zeta,x):$ $(\zeta,x,$ $\xib_{\eps}(\zeta,x))\in B,\;\eps \in (0,\eps_{B}]\}$ is bounded.
      \item  Let $B\subset D^{\iota}_{\eps}$ be a bounded set. Assume that $\phi\in C^{\infty}_{b}(B)$ and that $w$ satisfies \reff{eq: estimate varpi}. Then, there exists $\eps_{B}>0$ and $C_{B}>0$ such that
      $$
      |R_{\eps}(\zeta,x)| \le C_{B}(1+\eps|\xib_\eps|+\eps^2|\xib_\eps|^2)(\zeta,x)\;,
      $$
      for all $\eps \in (0,\eps_{B}]$ and $(\zeta,x)\in B$.
      \end{enumerate}
    \end{lemma}

\proof All over this proof, we work on $D_{\eps}^{\iota}$ and omit the argument for simplicity.

 \noindent   \step1{We first provide an expansion for   $\LSX \psi^{\eps}$.}
   The first term follows from the relation
   $x=\theta+\eps\xib_{\eps}$:
    \b*
    \LSX (v+\eps^{2}\phi)&=&\LStheta (v+\eps^{2}\phi)-\eps \mu \xib_{\eps} + \eps^{2}R_{1}^{\eps}\;,
    \e*
    with
    \b*
    R_{1}^{\eps}=  \eps \xib_{\eps} \left(   \mu\phi_{x} +\frac{\sigma^{2}}{2}\left( (2\theta  +\eps\xib_{\eps}) \phi_{xx}+2 s \phi_{xs}\right) \right).
    \e*
    Then, we use the fact that $\xib_{\eps}=\xib_{1}/\eps$ and the definitions of $\bar \sigma_{a}$ and $\bar \sigma_{\theta,a}$ in \reff{eq: def bar sigma a} and \reff{eq: def bar sigma theta a} to obtain
    \b*
    \LSX(\eps^{4}w^{\eps})
    &=&\frac{\eps^{2}}{2}(w_{\xi\xi}\circ\xib_{\eps})D\xib_{1}^{\top} \bar\sigma_{0}\bar \sigma_{0}^{\top} D\xib_{1}+\eps^{2}R_{2}^{\eps}
    \\
    &=&
    \frac{\eps^{2}}{2}(w_{\xi\xi}\circ\xib_{\eps})D\xib_{1}^{\top} \bar\sigma_{\theta,0}\bar \sigma_{\theta,0}^{\top} D\xib_{1}+\eps^{2}R_{3}^{\eps}\;,\\
    \e*
    where
    \b*
    R_{2}^{\eps}&=&  \eps^{2}(\LS w)\circ \xi_{\eps}   + \eps \left((w_{\xi}\circ\xib_{\eps}) \LSX\xib_{1}+  2s^{2}\sigma^{2} \partial_{s} \xib_{1} (w_{s\xi}\circ\xib_{\eps})\right)
    \e*
        and
    \b*
    R_{3}^{\eps} &=&R_{2}^{\eps}+
    \frac{\sigma^{2}}{2 }(w_{\xi\xi}\circ\xib_{\eps}) (D_{(s,x)}\xib_{1})^{\top} \left(\begin{array}{cc}0&s\eps\xib_{\eps} \\s\eps\xib_{\eps}&\theta\eps\xib_{\eps}+(\eps\xib_{\eps})^{2} \end{array}\right)D_{(s,x)}\xib_{1}\\
    &=&R_{2}^{\eps}+
   \eps\xib_{\eps} \frac{\sigma^{2}}{2}(w_{\xi\xi}\circ\xib_{\eps})(D_{(s,x)}\xib_{1})^{\top} \left(\begin{array}{cc}0&s  \\s &\theta+\eps\xib_{\eps} \end{array}\right)D_{(s,x)}\xib_{1}.
    \e*
  Combining the above expansions leads to
  \be\label{eq: expansion first term}
  \LSX \psi^{\eps}= \LStheta (v+\eps^{2}\phi)-\eps \mu \xib_{\eps}  +     \frac{\eps^{2}}{2}(w_{\xi\xi}\circ\xib_{\eps})D\xib_{1}^{\top} \bar\sigma_{\theta,0}\bar \sigma_{\theta,0}^{\top} D\xib_{1}+\eps^{2}(R_{1}^{\eps}+ R_{3}^{\eps}).
  \ee

 \noindent    \step2{We now focus on the operator $ \LPlSXh$ applied to $\psi^\eps$.}
    Since $\psi^\eps_{p}>0$ on $D^{\iota}_{\eps}$, we have
    \be\label{eq: proof expansion def a eps}
      \LPlSXh \psi^{\eps}= \LPlSX^{a^{\eps}} \psi^{\eps}  \;\mbox{ with }\;  a^\eps:=\frac{-\bar \sigma_{0}^{\top} D\psi^\eps}{\pi_p}
            \x\frac{1}{1 + \eps^2 \partial_{p}(\phi+\eps^{2}w^{\eps})/\pi_p}.
    \ee

    {\bf a. } We first provide an expansion for $a^{\eps}$   around $\hat a$ defined in \reff{eq: def hat a}. We start by performing a first order expansion on the right-hand side of  \reff{eq: proof expansion def a eps} to obtain
    \be\label{eq: first order taylor on a eps}
        a^\eps =\frac{-\bar \sigma_{0}^{\top} D\psi^\eps}{\pi_p}
            \x\left(1 -\eps^2 \partial_{p}(\phi+\eps^{2}w^{\eps})/\pi_p \right)
            +R_{4}^{\eps}\;,
    \ee
    where $R_{4}^{\eps}$ is a continuous map satisfying
    \b*
   | R_{4}^{\eps}|
    &\le&
   \frac{|\bar \sigma_{0}^{\top} D\psi^\eps| }{\pi_p} \frac{2}{(1+\iota)^{3}}\left|  \eps^2 \partial_{p}(\phi+\eps^{2}w^{\eps})/\pi_p \right|^{2} \;\;\mbox{  on $D^{\iota}_{\eps}$}
    \\
    &=&
    \frac{|\bar \sigma_{0}^{\top} D\psi^\eps| }{\pi_p} \frac{2}{(1+\iota)^{3}}\left| \eps^2\frac{\phi_p}{\pi_p}-\eps^3\frac{\theta_p(w_\xi\circ\xib_{\eps})}{\pi_p}+\eps^4\frac{(w_p\circ\xib_{\eps})}{\pi_p} \right|^{2}.
    \e*
Then, we obverse that
    \b*
    -\bar \sigma_{0}^{\top} D\psi^\eps&=& -\bar \sigma_{0}^{\top}Dv -\bar \sigma_{0}^{\top}D(\eps^{2} \phi+\eps^{4}w^{\eps})\\
        &=& -\bar \sigma_{\theta,0}^{\top}Dv+ \sigma \eps \xib_{\eps}-\bar \sigma_{\theta+\eps\xibu_{\eps},0}D(\eps^{2} \phi+\eps^{4}w^{\eps}).
    \e*
    By the definition of $\hat a$ in \reff{eq: def hat a}, dividing the above by $\pi_{p}=v_{p}$ implies
    \b*
    \frac{-\bar \sigma_{0}^{\top} D\psi^\eps}{\pi_p}=\hat a + \eps\frac{\sigma\xib_\eps}{\pi_p}     -   \frac{\bar \sigma_{\theta+\eps\xibu_{\eps},0}^{\top}D(\eps^{2} \phi+\eps^{4}w^{\eps})}{\pi_p}\;.
    \e*
    Recalling \reff{eq: first order taylor on a eps}, this leads to
    \be\label{eq: expansion a eps final}
     a^\eps& =&\hat a + \eps\frac{\sigma\xib_\eps}{\pi_p}     - \eps^2 \frac{\bar\sigma_{\theta,\hat a}^{\top}D\phi}{\pi_p}
                + R_{5}^{\eps}\;,
    \\
    \label{eq: expansion a eps squared final}
            (a^\eps)^2 &=&\left(\hat a\right)^2+2\eps \hat a\frac{\sigma\xib_\eps}{\pi_p}
            + \eps^2
                \left[
                    \left( \frac{\sigma\xib_\eps}{\pi_p} \right)^2  -2\hat a \bar \sigma_{\theta,\hat a}^{\top}D\phi/\pi_p
                \right]
            +R_{6}^{\eps}\;,
    \ee
               where
    \b*
     R_{5}^{\eps}&:=&    R_{4}^{\eps}-\eps^{2}\left[\eps\xib_{\eps}\frac{\sigma\phi_{p}}{ (\pi_{p})^{2} }+\frac{\bar\sigma_{\eps\xibu_{\eps},0}^{\top}D\phi}{\pi_{p}}\right]+\eps^{4}\frac{w^{\eps}_{p}}{ \pi_{p} }\left( - \hat a   -  \eps\xib_{\eps}\frac{\sigma}{\pi_{p}}+\eps^{2} \frac{\bar \sigma_{\theta+\eps\xibu_{\eps},0}^{\top}D\phi}{\pi_{p}}\right)\\
     &&-\eps^{4} \frac{\bar \sigma_{\theta+\eps\xibu_{\eps},0}^{\top}Dw^{\eps}}{ \pi_p }\left(1 -\eps^2 \partial_{p}(\phi+\eps^{2}w^{\eps})/\pi_p \right)
     +\eps^{4}\frac{\phi_{p}}{\pi_{p}^{2}} \bar\sigma_{\theta+\eps\xibu_{\eps},0}^{\top}D\phi\;,
    \\
   R_{6}^{\eps}&=&\left(- \eps^2 \frac{\bar \sigma_{\theta,\hat a}^{\top}D\phi}{\pi_p}
                + R_{5}^\eps  \right)^{2}+2\hat a R_{5}^{\eps}+2\frac{\sigma\xib_\eps}{\pi_p}  \left(- \eps^2 \frac{\bar \sigma_{\theta,\hat a}^{\top}D\phi}{\pi_p}
                + R_{5}^\eps  \right)\;.
   \e*

   {\bf b.} We now plug the expansions \reff{eq: expansion a eps final} and \reff{eq: expansion a eps squared final} in  the left-hand side equality in \reff{eq: proof expansion def a eps} to obtain
           \be
        \LPlSXh\psi^\eps &=&
            \LPlStheta^{\hat a} v
            +\eps\left[
                \pi_{pp}\hat a\frac{\sigma\xib_\eps}{\pi_p}
                + \sigma^{2} s\pi_{sp}\frac{\xib_\eps}{\pi_p}
            \right]  \label{eq: expansion LSlSXh}\\
        && +\eps^2 \left( \frac12\pi_{pp}\left[
                            \left( \frac{\sigma\xib_\eps}{\pi_p} \right)^2  -2\hat a \left( \frac{\bar \sigma_{\theta,\hat a}^{\top}D\phi}{\pi_p}\right)
                        \right]
                        - \sigma s\pi_{sp}\frac{\bar \sigma_{\theta,\hat a}^{\top}D\phi}{\pi_p}+\LPlStheta^{\hat a }\phi
 \right)\nonumber
 	\\
	&& +\eps^2\left(\frac12 (w_{\xi\xi}\circ\xib_{\eps} )D\xib_{1}^{\top}(\bar \sigma_{\theta,\hat a}\bar \sigma_{\theta,\hat a}^{\top}-\bar \sigma_{\theta,0}\bar \sigma_{\theta,0}^{\top})D\xib_{1} + R_{7}^{\eps}\right)\;, \nonumber
    \ee
    with
    \b*
        R_{7}^{\eps}&=&\frac12 R_{6}^{\eps}\pi_{pp}+R_{5}^{\eps}\sigma s \pi_{sp}
        +\frac12 ((a^{\eps})^{2}-(\hat a)^{2})(\eps^{2}\phi_{pp}+\eps^{4}w^{\eps}_{pp}) +\sigma s(a^{\eps}-\hat a)(\eps^{2} \phi_{sp}+\eps^{4}w^{\eps}_{sp})
        \\
        &&+ [(a^{\eps}-\hat a)(\eps \xib_{\eps}+\theta)+\hat a\eps \xib_{\eps}]\sigma \phi_{px}
        \\
        &&+\frac{\eps^{3}}{2}(\hat a)^{2}\left(\eps w_{pp}-2\theta_{p} w_{p\xi} -\theta_{pp}w_{\xi} \right)\circ\xib_{\eps}\\
        && +\eps^{3} \hat a\sigma\left(\eps s w_{sp} -s \theta_{p} w_{s\xi}-s\theta_{s}w_{p\xi}-s\theta_{sp} w_{\xi}+\theta w_{p\xi}\right)\circ\xib_{\eps} .
        \e*

   \noindent \step3{It remains to combine the results of Steps 1 and 2.}          We first observe that \reff{eq: PDE pi} and the definition     $\hat a$ implies that
       \b*
        \LStheta v+ \LPlStheta^{\hat a} v=\LStheta v+ \LPlSthetah v=0.
        \e*
        Second, we use  \reff{eq: def hat a} and the identity $v=\pi-x$ to obtain
        $$
        \hat a=\frac{\frac{\mu}{\sigma}\pi_{p}- \sigma s\pi_{ps}}{\pi_{pp}},
        $$
        which
        leads to
        $$
        \eps\xib_{\eps}\left(-\mu+\pi_{pp}\hat a\frac{\sigma}{\pi_p}
                + \sigma^{2} s\pi_{sp}\frac{1}{\pi_p}\right)=0,
        $$
        and
        \b*
        \LStheta \phi+\LPlStheta^{\hat a}\phi-  \pi_{pp}  \hat a \left( \frac{\bar \sigma_{\theta,\hat a}^{\top}D\phi}{\pi_p}\right)
                        - \sigma s\pi_{sp}\frac{\bar \sigma_{\theta,\hat a}^{\top}D\phi}{\pi_p}
                     &=&
           \LStheta \phi+\LPlStheta^{\hat a}\phi  -\frac{\mu}{\sigma} \bar \sigma_{\theta,\hat a}^{\top} D\phi   \\
           &=& (\Hc+\LXlSP^{\hat a})\phi.
        \e*
        Finally, we use the identities  $\xib_{1}=\theta-x$ and   $\hat a=(\theta-s\pi_{s})\sigma/\pi_{p}$, recall \reff{eq: relation thetat with hat a},
        to obtain
        \b*
          \sigma^{2}\delta^{2}&=&  D\xib_{1}^{\top} \bar\sigma_{\theta,0}\bar \sigma_{\theta,0}^{\top} D\xib_{1}+D\xib_{1}^{\top}(\bar \sigma_{\theta,\hat a}\bar \sigma_{\theta,\hat a}^{\top}-\bar \sigma_{\theta,0}\bar \sigma_{\theta,0}^{\top})D\xib_{1},
         \e*
         where $\delta$ is defined in \reff{eq: def delta}.
         The above identities  combined with \reff{eq: expansion first term} and \reff{eq: expansion LSlSXh} leads to \reff{eq: lem expansion} for $R_{\eps}$ defined as
         \be\label{eq : def R eps iota}
         R_{\eps}:= R_{1}^{\eps}+ R_{3}^{\eps}+R_{7}^{\eps}.
         \ee
    \noindent\step4{} The estimates on $R_{\eps}$ follow from direct computations.
         \ep

    \subsection{Viscosity subsolution property}

     \begin{proposition}\label{prop : subsol u*} Let the conditions of Theorem \ref{thm: main} hold. Then, $u^{*}$ is a viscosity subsolution of \reff{eq: 2nde corrector equation}.
     \end{proposition}

      \noindent  \textbf{Proof.}
        Let $\zeta_{o}\in\Di$ and $\vp\in C^{1,2}(\Di)$ be such that
        $$
            \max_{\Di}(\mbox{strict})(u^*-\vp)=(u^*-\vp)(\zeta_o).
        $$
        By Lemma \ref{lem: u does not depend on x},
        there exists $(\zeta^\eps)_{\eps>0}$ satisfying
        \beq\label{eq: sous sol in domain convergence t, s, p_eps}\begin{matrix}
            \displaystyle \zeta^\eps\underset{\eps\downarrow0}{\longrightarrow}\zeta_o,
            \qquad x^\eps:=\theta(\zeta^\eps)\underset{\eps\downarrow0}{\longrightarrow}\theta(\zeta_o)=:x_o,\\
            \displaystyle
            u^{\eps*}(\zeta^\eps,x^\eps)\underset{\eps\downarrow0}{\longrightarrow}u^*(\zeta_o)
            \mbox{ and }\Delta_\eps := u^{\eps*}(\zeta^\eps,x^\eps)-\vp(\zeta^\eps)
            \underset{\eps\downarrow0}{\longrightarrow}0.
        \end{matrix}\eeq
                Assumptions \ref{ass: u eps finite} and  \ref{ass: pi smooth} entail the existence of $\bar r_o>0$, $0<r_{o}\le \bar r_{o}$ and $\eps_o>0$
        such that
        $$
            \bar m := \sup \left\{ u^{\eps\ast}(\zeta,x), (\zeta,x)\in B_o, \eps\in(0,\eps_o] \right\}<\infty,
        $$
       and
        \beq\label{eq: subsol Ac theta bounded r_o}
            \theta \in\bar B_{\frac{\bar r_{o}}{4}}(x_o)\quad\mbox{ on }\quad \bar B_{r_{o}}(\zeta_o),
        \eeq
        where $B_{o}:=B_{r_{o}}(\zeta_o)\times B_{\bar r_{o}}(x_o)$.
        After possibly changing $\eps_o$, we can also assume that
        \beq\label{eq: subsol Ac eps_1}
            \left|\zeta^\eps-\zeta_o\right| \vee
                        \left|x^\eps-x_o\right|\le \frac{r_{o}}{4}
            \quad\mbox{and}\quad
                \left|\Delta_\eps\right|\le1
            \quad\pourtout \eps\in(0,\eps_o].
        \eeq
       {We have}
        \beq\label{eq: subsol Ac sup for u eps}
            u^{\eps *}\le\bar m \qquad\mbox{ on}\;\;  B_o\quad\mbox{  for}\quad\eps\in(0,\eps_o],
        \eeq
        and, by Assumption \ref{ass: pi smooth},
        \beq\label{eq: v_p greater iota}
            \pi_{pp}\wedge\pi_p >\iota \;\mbox{on}\; B_{r_{o}}(\zeta_o),\;\mbox{ for some $\iota\in (0,1)$.}
        \eeq

        \step1{We first construct a suitable test function for $v^{\eps}$, for $\eps\in (0,\eps_{o}]$.}

        Since the function $\vp$ is continuous, 
        $$
            \sup \left\{ 2+\bar m - \vp(\zeta)\; ; \; \zeta\in \bar B_{r_{o}}(\zeta_o) \right\}=: \bar M<+\infty.
        $$
        On the other hand, \reff{eq: subsol Ac eps_1} implies that  there is $\gamma>0$ such that
        \be\label{eq: dist bigger gamma}
              \left| \zeta-\zeta^\eps \right|^4\ge \gamma
                \quad\mbox{for}\quad  \zeta \in\bar B_{r_{o}}(\zeta_o)\backslash\bar B_{\frac{r_{o}}2}(\zeta_o).
        \ee
        We   choose a strictly non-negative constant $c_o$  satisfying $c_o(\gamma\wedge(\frac{r_o}4)^4)\ge \bar M$
        and   define for   $\eps\in(0,1)$
        $$
            \phi^\eps:(\zeta,x)\in \D\x \R\mapsto c_o \left(
                        \left| \zeta-\zeta^\eps \right|^4
                        + \left| x-\theta(\zeta) \right|^4
                    \right).
        $$
        Consider now the following subset of $\bar B_o$:
        $$
            B_{o,\frac12} :=
                \left\{
                    (\zeta,x)\in \bar B_o \st  \zeta \in \bar B_{\frac{r_{o}}{2}}(\zeta_o) \mbox{ and }x\in\bar B_{\frac{\bar r_o}2}(x_o)
                \right\}.
        $$
        It follows from \reff{eq: dist bigger gamma}, \reff{eq: subsol Ac theta bounded r_o} and the choice of $c_o$ that
        \beq\label{eq: phi eps grand}
             \displaystyle\phi^\eps  \ge 2+\bar m-\vp
                \quad \mbox{ on }  \bar B_o\backslash B_{o,\frac12}.
        \eeq
        We now define, for  $\eta\in(0,1]$,
        $$
            \psi^{\eps,\eta}:=
                v
                + \eps^2
                    \left( \Delta_\eps + \vp + \phi^\eps  
                    \right)
                + \eps^4 (1+\eta)\varpi\circ\xib_\eps,
        $$
        where the function $\xib_{\eps}$ is defined in \reff{eq: def xi} and $\varpi$ is given in Lemma \ref{lem: existence and property varpi}.

       \step2{  Given $\eps\in (0,   \eps_{o}]$ and $\eta \in (0,1]$, we now show that  $v^{\eps*}-\psi^{\eps,\eta}$       admits a local maximizer $(\tilde \zeta^\eps,\tilde x^\eps)$ in $B_o$. }

       Note that, a-priori, this local maximizer should depend on $\eta$. We shall not emphasize this to alleviate notations but will come back to this point at the end of the proof.
        We set
        $$\bal
            I^{\eps,\eta}
            &:=
                 u^{\eps *}
                - \Delta_\eps
                - \vp - \phi^\eps
                - \eps^2 (1+\eta)\varpi\circ\xib_\eps.
        \eal$$
        Combining
        the fact that $\varpi(\cdot,0)=0$, see Lemma \ref{lem: existence and property varpi}, \reff{eq: subsol Ac eps_1} and the
        definitions of $x^{\eps}$, $\Delta_\eps$ and $\phi^\eps$, we obtain
        $$
            \sup_{\bar B_o}I^{\eps,\eta}\ge \sup_{\bar B_{o,\frac12}}I^{\eps,\eta}
                \ge I^{\eps,\eta}(\zeta^\eps,x^\eps)=0.
        $$
        On the other hand, by \reff{eq: subsol Ac eps_1},  \reff{eq: subsol Ac sup for u eps}, \reff{eq: phi eps grand}, the fact that $\varpi\ge0$, see Lemma \ref{lem: existence and property varpi}, and the defnition of $\bar m$, we have
        $$
              I^{\eps,\eta}
                \le   u^{\eps *} -\bar m-1- \eps^2(1+\eta) \varpi\circ\xib_\eps<0\lpt7
             \quad  \mbox{ on }\quad \bar B_o\backslash \bar B_{o,\frac12},
         $$
         after possibly changing $\eps_{0}$.
    Also 
        $I^{\eps,\eta}$ is upper-semicontinuous.  Hence,
        we may find a maximizer
        $(\tilde \zeta^\eps,\tilde x^{\eps})\in \bar B_{o,\frac12}\subset B_o$
         which  satisfies
        \be\label{eq: bound sequence t s p x eps}
            I^{\eps,\eta}\left(\tilde \zeta^\eps,\tilde x^{\eps}\right)\ge0
            \quad\mbox{and}\quad
            \left|\eps\xib_\eps(\tilde \zeta^\eps,\tilde x^\eps)\right|
                \vee\left| \tilde \zeta^\eps - \zeta_o \right|
                \le r_1\;,
        \ee
        for some constant $r_1>0$. We recall that
     \be\label{eq: def tilde xi eps eps}
 \eps   \xib_\eps(\tilde \zeta^\eps,\tilde x^\eps)=  \tilde x^\eps-\theta(\tilde \zeta^\eps).
     \ee

        \step3
        {
        We now prove that there exists $\bar\eps_o\le\eps_o$ such that for all $\eps\in(0,\bar\eps_o]$
        we have
        \beq\label{eq: subsol xi eps bounded}
         - \hat\xib(\tilde \zeta^\eps)  < \xib_\eps(\tilde \zeta^\eps,\tilde x^\eps)< \hat  \xib(\tilde \zeta^\eps),
        \eeq
        where $\hat \xib$ is given in Lemma \ref{lem: existence and property varpi}.
        }

We only prove the right hand-side. The other inequality is proved similarly.
        We first observe that  Theorem \ref{thm: pde veps} and  step 2 imply that
        \beq\label{eq: dpe subsol v eps boundary}
                    -\eps^3+1+\psi^{\eps,\eta}_x\left(\tilde \zeta^\eps,\tilde x^{\eps}\right)\le0.
        \eeq
        Recalling the definitions of $\psi^{\eps,\eta}, v$ and $\phi^\eps$,
        direct computations lead to
        $$
            1+\psi^{\eps,\eta}_x\left(\tilde \zeta^{\eps},\tilde x^{\eps}\right)
                =4\eps^2 c_o \left(  \eps\xib_\eps(\tilde \zeta^\eps,\tilde x^\eps) \right)^3
                    +\eps^3(1+\eta)\varpi_\xi\circ\xib_{\eps}(\tilde \zeta^{\eps}, \tilde x^\eps ),
        $$
        so that we may rewrite
        \reff{eq: dpe subsol v eps boundary} as
        \be
          -\eps
                + \eps(1+\eta)\varpi_\xi\circ\xib_{\eps}(\tilde \zeta^{\eps}, \tilde x^\eps )
                \le-4c_o\left( \eps\xib_\eps(\tilde \zeta^\eps,\tilde x^\eps) \right)^3.&\label{eq: sous sol 1st order 1 inside domain}
        \ee
        Assume now that the right hand-side of  \reff{eq: subsol xi eps bounded} does not hold for all $\eps>0$, small enough.
        Then, there exists a sequence $(\eps_k)_{k\ge1}$
        satisfying $\eps_k\rightarrow0$ as $k\rightarrow\infty$         such that
        $$
             \xib_{\eps_{k}}(\tilde \zeta^{\eps_{k}},\tilde x^{\eps_{k}})
                \ge\hat  \xib\left(\tilde \zeta^{\eps_k}\right).
        $$
        Recall from Lemma \ref{lem: existence and property varpi}   that this implies that
        $$
        \varpi_\xi\circ \xib_{\eps_{k}} (\tilde \zeta^{\eps_k},\tilde x^{\eps_k})=1\;\mbox{ and }\;  \xib_{\eps_{k}} (\tilde \zeta^{\eps_k},\tilde x^{\eps_k}) > 0.
        $$
        Combined with \reff{eq: sous sol 1st order 1 inside domain} the later leads to a contradiction since $c_{o},\eta,\eps_{k}>0$.

        \step4{  We now prove that there is $\bar \xi\in\R$ such that
        \beq\label{eq: subsol intermed form}
            0\ge\left(-\frac12\frac{\pi_{pp}}{\left(\pi_p\right)^2}\sigma^2\bar \xi^2
            -\Hc\vp - \frac12\sigma^2\delta^2(1+\eta)\varpi_{\xi\xi}(\cdot,\bar  \xi)\right)
            \left(\zeta_o\right).
        \eeq}
 Recall that $\hat \xib$ is a continuous functions. In view of \reff{eq: subsol xi eps bounded} and \reff{eq: bound sequence t s p x eps}, it follows that
 \be\label{eq: bound t x s p xi eps}
 (\tilde \zeta^\eps,\tilde x^{\eps},  \xib_{\eps}(\tilde \zeta^\eps,\tilde x^{\eps}))_{0<\eps\le \bar \eps_{o}} \;\mbox{ is bounded}.
 \ee
We can then find a  sequence $(\eps_n)_{n\ge1}\subset(0,\bar\eps_o]$ such that
        $\eps_n\rightarrow0$ as $n\rightarrow\infty$ and
        \be\label{eq: lim seq exists}
        (\tilde \zeta^{\eps_n},\tilde x^{\eps_n}, \xib_{\eps_{n}}(\tilde \zeta^{\eps_n},\tilde x^{\eps_n}))\rightarrow(\bar  \zeta,\bar  x,\bar  \xi)\in \D\x \R \x \R\; \mbox{ as } n\to \infty.
        \ee
        Moreover, classical arguments show that
        \be\label{eq: lim is the good one }
        (\bar  \zeta,\bar  x)=(\zeta_o,x_o).
        \ee
        Observe for later use that
        \be\label{eq: hat xi bounded}
        -\hat \xib(\zeta_o)\le \bar  \xi\le\hat \xib(\zeta_o)\;,
        \ee
        by \reff{eq: subsol xi eps bounded} and the continuity of $\hat \xib$.
        By Step 2 and Theorem \ref{thm: pde veps} again,  we have
        \beq\label{eq: dpe subsol v eps hold}
                \left\{- \LSX\psi^{\eps_n,\eta} - \LPlSXh \psi^{\eps_n,\eta}\right\}
                    \left( \tilde \zeta^{\eps_n},\tilde x^{\eps_n} \right)
            \le0
            \quad\pourtout n\ge1.
        \eeq
        Moreover, \reff{eq: bound t x s p xi eps}, {\reff{eq: v_p greater iota}} and Lemma \ref{lem: existence and property varpi} imply that we can apply Lemma \ref{lem: remainder estimate} to $\psi^{\eps_{n},\eta}$. For $n$ large enough:
        \b*
            0\ge\left( -\frac{\pi_{pp}}{2\left(\pi_p\right)^2}\sigma^2\xib_{\eps_{n}}^2
                    -\Hc\bar\vp^{\eps_n} - \LXlSP^{\hat a} \phi^{\eps_n} - \frac{\sigma^2\delta^2(1+\eta)(\varpi_{\xi\xi}\circ\xib_{\eps_{n}})}2 +R_{\eps_{n}}
                \right)
                \left(\tilde \zeta^{\eps_n},\tilde x^{\eps_n} \right),
        \e*
        where
        $$
        \bar\vp^{\eps_n} :=\Delta_{\eps_n} +\vp + \phi^{\eps_n},
        $$
        and  $R_{\eps_{n}} (\tilde \zeta^{\eps_n},\tilde x^{\eps_n} )\to 0$ as $n\to \infty$.          Sending $n\rightarrow\infty$ and using \reff{eq: def tilde xi eps eps}, \reff{eq: lim seq exists} and \reff{eq: lim is the good one } provides
        \reff{eq: subsol intermed form}.

        \step5{We can now conclude the proof.}          By the construction of $\varpi$ as a solution of the first corrector equation \reff{eq: first corrector equation}
        and by \reff{eq: hat xi bounded},  we have
        $$
            \frac12(\frac{\pi_{pp}}{\left(\pi_p\right)^2}\sigma^{2})(\zeta_o) \bar  \xi^2
                + \frac12(\sigma^2\delta^{2})(\zeta_o)\varpi_{\xi\xi}\left(\zeta_o,\bar  \xi\right)=h(\zeta_o)\;,
        $$
        which plugged into \reff{eq: subsol intermed form} gives
        \beq\label{eq: subsol with eta}
            -\Hc\vp(\zeta_o) \le h(\zeta_o)
            +\eta\frac12\sigma^2\delta(\zeta_o)^2\varpi_{\xi\xi}\left(\zeta_o,\bar \xi\right).
        \eeq
         Finally we note that, although $\bar  \xi$ as constructed in Step 4 above depends on $\eta$,  $\zeta_o$ does not depend on this parameter, and therefore $|\varpi_{\xi\xi}(\zeta_{o},\cdot)|$ is bounded by Lemma \ref{lem: existence and property varpi}. Sending $\eta\to 0$ in the above inequality leads to
        \b*
            -\Hc\vp(\zeta_o) \le h(\zeta_o).
        \e*

        \qed


\subsection{Viscosity supersolution property}

\def\hf{{\mathfrak h}}

         For sake of completeness, we report here \cite[Lemma 5.4]{PoSoTo12} that will be used in the proof below.
        \begin{lemma}\label{lem: construction of h eta}
          For all $\eta\in(0,1)$,
          there exists $c_\eta >1$ and a smooth function $\hf_{\eta} :\R\rightarrow[0,1]$ satisfying $\hf_\eta=1$ on $[-1,1]$,
          $\hf_\eta=0$ on $[-c_\eta,c_\eta]^c$ and
          \beq\label{eq: estimate h eta}
            |x| |\hf_\eta'(x)|\le  \eta,
            \qquad\mbox{and}\qquad
            |x| |\hf_\eta''(x)|\le 2C^*,
          \eeq
          for some constant $C^*>0$ independent of $\eta$.
        \end{lemma}

     \begin{proposition}\label{prop : supersol u*} Let the conditions of Theorem \ref{thm: main} hold. Then, $u_{*}$ is a viscosity supersolution of \reff{eq: 2nde corrector equation}.
     \end{proposition}

      \noindent  \textbf{Proof.}
        Let $\zeta_o\in \Di$ and $\vp\in C^{1,2}(\Di)$ be such that
        $$
            \min_{\Di}(\mbox{strict})(u_*-\vp)=(u_*-\vp)(\zeta_o)=0.
        $$
        By Lemma \ref{lem: u does not depend on x} and the continuity of $\vp$,
        there exists $(\zeta^\eps)_{\eps>0}$ such that
        \beq\label{eq: sur sol convergence t, s, p_eps}\begin{matrix}
            \displaystyle \zeta^\eps \underset{\eps\downarrow0}{\longrightarrow} \zeta_o,
            \qquad x^\eps:=\theta(\zeta^\eps)\underset{\eps\downarrow0}{\longrightarrow}\theta(\zeta_o)=:x_o\;,\\
            \displaystyle u^\eps_{*}(\zeta^\eps,x^\eps)\underset{\eps\downarrow0}{\longrightarrow}u_*(\zeta_o)
            \mbox{ and } \Delta_\eps := u^\eps_{*}(\zeta^\eps,x^\eps)-\vp(\zeta^\eps)
            \underset{\eps\downarrow0}{\longrightarrow}0\;.
        \end{matrix}\eeq
        Let $r_o>0$ and $\eps_o\in(0,1]$ be such that
            \beq\label{eq: supersol Ac t eps small enough}
                |\zeta^\eps-\zeta_o|\le \frac{r_o}{2}
                \quad\mbox{and}\quad |\Delta_\eps|\le 1
                \quad\pourtout\eps\le\eps_0.
            \eeq

         \step1{We fix $\eps \in (0,\eps_{o}]$ and   construct a  first test function for $u^{\eps}_{*}$.  }

            Since $\vp$ is smooth, there exists a constant $M<\infty$ such that
            \beq\label{eq: supersol Ac sup vp finite c_o}
                \sup\left\{\vp(\zeta)\; ; \; \zeta\in\bar B_{r_o}(\zeta_o)\right\} \le M-4.
            \eeq
            By \reff{eq: supersol Ac t eps small enough}, there exists a finite $d>0$ such that
            $|\zeta-\zeta^\eps|^4\ge d$ for all $\zeta\in\partial B_{r_o}(\zeta_o)$. We fix
 $c_o>0$ such that $c_od\ge M$ and  define
            $$
                \phi^\eps(\zeta) := \vp(\zeta) + \Delta_{\eps} - c_o\left( |\zeta-\zeta^\eps|^4\right).
            $$
            It follows  from \reff{eq: supersol Ac t eps small enough},
            \reff{eq: supersol Ac sup vp finite c_o} and the choice of $c_o$ that
            \beq\label{eq: supersol Ac diff vp and phi eps}
                    -\phi^\eps
                    \ge 3\quad\mbox{ on }\quad \partial B_{r_o}(\zeta_o)
                    .
            \eeq
            Observe for later use that
            \beq\label{eq: supersol Ac u - vp = 0 in t eps}
                (u^\eps_{*}-\phi^\eps)(\zeta^\eps,x^\eps)=0\;,
            \eeq
		by the definition of $\Delta_{\eps}$.

               For $\eta\in(0,1)$, we now set
            $$
                \psi^{\eps,\eta}  := v
                    + \eps^2\phi^\eps
                    + \eps^4 \left( 1-\eta \right) (\varpi H_{\eta})\circ \xib_\eps\;,
            $$
            in which
            $$
            H_{\eta}:\xi \in \R\mapsto \hf_\eta\left(\frac\xi{\xi_{*}}\right)\;,
            $$
            for some $\xi_{*}\ge 1$ to be chosen later on, see \reff{eq: def xi star} in Step 6, and where $\hf_\eta$ is as in  Lemma \ref{lem: construction of h eta}.

            \step2{Let $\Qc_o:=\bar B_{r_o}(\zeta_o)\x \R$ and fix $\eps\in (0,\eps_{o}]$. We now show that, for each $n\ge 1$,  there exists $(\hat \zeta^{\eps,n},\hat x^{\eps,n})\in$ {\rm Int}$(\Qc_o)$ satisfying
            \beq\label{eq: supersol Ac n-1 optimal minimizer for I}
                I^{\eps,\eta}\left(\hat \zeta^{\eps,n},\hat x^{\eps,n}\right)\le \inf_{\Qc_o}I^{\eps,\eta} +\frac1{2n},
            \eeq
            in which
            \be
                 I^{\eps,\eta} &:=&\eps^{-2}\left(v^{\eps}_{*}-\psi^{\eps,\eta}\right)
               =   u^{\eps}_{*}-\phi^\eps - \eps^2(1-\eta)(\varpi H_{\eta})\circ\xib_{\eps}. \label{eq: supersol Ac I eps eta with bar u}
            \ee
            }
            Note that $\xib_{\eps}(\zeta^{\eps},x^{\eps})=0$ since $x^\eps=\theta(\zeta^\eps)$. Recalling that  $\varpi(\cdot,0)=0$ by Lemma \ref{lem: existence and property varpi},   \reff{eq: supersol Ac u - vp = 0 in t eps} implies that
            \beq\label{eq: supersol Ac I=0 in t eps}
                I^{\eps,\eta}(\zeta^\eps,x^\eps)=0.
            \eeq
            On the other hand, \reff{eq: supersol Ac I eps eta with bar u} combined with Remark \ref{rem: v le veps}, Remark \ref{rem: property varpi} and Lemma \ref{lem: construction of h eta}  implies that
            \b*
                I^{\eps,\eta} &\ge& -\phi^\eps-\eps^2(1-\eta)\{|\xib_\eps|\1_{\{|\xibu_{\eps}|\le c_\eta \xi_{*}\}}\}
                  \ge -\phi^\eps -\eps^2(1-\eta)c_\eta\xi_{*}.
            \e*
            In particular,
            \be
            I^{\eps,\eta}\ge -\phi^\eps -1\qquad\mbox{ if }\;  \eps\le\eps_\eta:=  \eps_o\wedge((1-\eta)c_\eta\xi_{*})^{-\frac12}.\label{eq: supersol Ac min for I eps eta}
            \ee
           The set $\bar B_{r_o}(\zeta_o)$ being compact, the inf over $\bar B_{r_{o}}(\zeta_{o})$ of the right-hand side is finite, which    proves our claim.

        \step3{For   $\eta\in(0,1), \eps\in(0,\eps_\eta]$ and $n\ge1$, we now construct a $C^{2}$ function $\psi^{\eps,\eta,n}$ and $(\zeta^{\eps,n},x^{\eps,n})\in${\rm Int}$(\Qc_o)$ such that
        $$
            \min_{ \Qc_o}(v^\eps-\psi^{\eps,\eta,n})=(v^\eps-\psi^{\eps,\eta,n})(\zeta^{\eps,n},x^{\eps,n}).
        $$}
           Let $f\in C^\infty_{b}(\R)$ be an even function satisfying
            $0\le f\le1$, $f(0)=1$ and $f(x)=0$ whenever $|x|\ge1$.
            We set
            $$
                \psi^{\eps,\eta,n}(\cdot,x) := \psi^{\eps,\eta}(\cdot,x)+\frac{\eps^2}n f\left(  x-\hat x^{\eps,n}  \right)
            $$
            and
            $$
                I^{\eps,\eta,n}(\cdot,x) :=\frac1{\eps^2}\left({v^\eps_{*}}-\psi^{\eps,\eta,n}\right)(\cdot,x)
                    =I^{\eps,\eta}(\cdot,x)-\frac1nf\left(  x-\hat x^{\eps,n}  \right).
            $$
            By \reff{eq: supersol Ac n-1 optimal minimizer for I} and the identity $f(0)=1$,
            \beq\label{eq: link I eps eta and I eps eta n}
                I^{\eps,\eta,n}\left(\hat \zeta^{\eps,n},\hat x^{\eps,n}\right)
                    =I^{\eps,\eta}\left(\hat \zeta^{\eps,n},\hat x^{\eps,n}\right)-\frac1n
                    \le \inf_{ \Qc_o}I^{\eps,\eta} -\frac1{2n}.
            \eeq
            Moreover, by the definition of $f$,
            $$
                I^{\eps,\eta,n} =I^{\eps,\eta}\;\mbox{ on } \Qc_o\backslash\Qc_1^n,
            \;\mbox{ where }\;
            \Qc^n_1:=\{(\zeta,x)\in\Qc_o\st|x-\hat x^{\eps,n}|\le1\}.
            $$
            Since $(\hat \zeta^{\eps,n}, \hat x^{\eps,n})\in\Qc^n_1$, the later combined with   \reff{eq: link I eps eta and I eps eta n}
             implies that
            $$
                \inf_{\Qc_1^n}I^{\eps,\eta,n}<\inf_{\Qc_o}I^{\eps,\eta}
                    \le\inf_{\Qc_o\backslash\Qc_1^n}I^{\eps,\eta}=\inf_{\Qc_o\backslash\Qc_1^n}I^{\eps,\eta,n},
            $$
            so that
            $$
                \inf_{\Qc_o}I^{\eps,\eta,n}=\inf_{\Qc_1^n}I^{\eps,\eta,n}.
            $$
            By the lower semi-continuity of $I^{\eps,\eta,n}$
            and the compactness of $\Qc_1^n$,
            we can then find $(\zeta^{\eps,n},x^{\eps,n})\in\Qc_o$ which minimizes $I^{\eps,\eta,n}$ on $\Qc_o$.
            It remains to show that it belongs to {\rm Int}$(\Qc_o)$. Indeed,  the left hand-side of \reff{eq: supersol Ac t eps small enough}, the property  $f\ge0$, and  \reff{eq: supersol Ac I=0 in t eps} imply that
            $$
                I^{\eps,\eta,n}\left(\zeta^{\eps,n},x^{\eps,n}\right)\le I^{\eps,\eta,n}\left(\zeta^\eps,x^\eps\right)
                    \le I^{\eps,\eta}\left(\zeta^\eps,x^\eps\right)=0,
            $$
            whereas by \reff{eq: supersol Ac diff vp and phi eps},
            \reff{eq: supersol Ac min for I eps eta} and the fact that $-f\ge-1$, we have
            \beq\label{eq: supersol Ac (t,s,p) on the boundary of the ball}
                \displaystyle
                I^{\eps,\eta,n}
                \ge I^{\eps,\eta} -\frac1n
                \ge 2-\frac1n>0
                \qquad\mbox{on } \partial\Qc_o={\partial   B_{r_{o}}(\zeta_{o})}\x \R.
            \eeq

        \step4{Given $\eta\in (0,1)$ and $\eps \in (0,\eps_{\eta}]$, we now show that there exists $N_{\eps,\eta}\ge 1$ such that
        \beq\label{eq: supersol Ac spersol for psi eps eta n}
            -\left(\LSX v^\eps+\LPlSXh \right)\psi^{\eps,\eta,n}\left(\zeta^{\eps,n},x^{\eps,n}\right)\ge0\;\mbox{ for } n\ge N_{\eps,\eta}.
        \eeq}
In view of step 3 and Theorem \ref{thm: pde veps},
            it suffices to show that
            $$
                \max\left\{
                    -\eps^3+1+\psi^{\eps,\eta,n}_x;
                    -\eps^3-(1+\psi^{\eps,\eta,n}_x)
                \right\}\left(\zeta^{\eps,n},x^{\eps,n}\right)<0,
            $$
            or equivalently that
            $$
                \left| 1+\psi^{\eps,\eta,n}_x \right|\left(\zeta^{\eps,n},x^{\eps,n}\right)<\eps^3.
            $$
            Recalling that $f\in C^\infty_{b}(\R)$ is even,
            we first compute
            \b*
                1+\psi^{\eps,\eta,n}_x\left(\zeta^{\eps,n},x^{\eps,n}\right)& =&
                    \eps^3(1-\eta)(\varpi H_{\eta})_\xi\circ\xib_{\eps}\left(\zeta^{\eps,n},x^{\eps,n}\right)
                     + \frac{\eps^2}n   f'\left( {|}x^{\eps,n}-{\hat x^{\eps,n}}{|}\right).
            \e*
           Since $f\in C^{\infty}_{b}(\R)$ is constant outside $[-1,1]$, there exists $0<c_f<+\infty$,
            which does not depend on $\eps$ nor $n$, such that
            $$
                \left|1+\psi^{\eps,\eta,n}_x\left(\zeta^{\eps,n},x^{\eps,n}\right)\right| =
                    \eps^3(1-\eta)
                        \left(\left| \varpi_\xi H_{\eta} \right| + \left| \varpi H_{\eta}' \right|\right)\circ\xib_{\eps}
                            \left(\zeta^{\eps,n},x^{\eps,n}\right)
                            + \frac{\eps^2c_f}n.
            $$
            In view of \reff{eq: first corrector equation}, (ii) of Lemma \ref{lem: existence and property varpi}, Remark \ref{rem: property varpi} and the fact that $|H_{\eta}|\le 1$ {by Lemma \ref{lem: construction of h eta}}, this implies that
            $$
                \left|1+\psi^{\eps,\eta,n}_x\left(\zeta^{\eps,n},x^{\eps,n}\right)\right| =
                    \eps^3(1-\eta)
                        \left(1+ \frac{|\xib_{\eps}|}{\xi_{*}}\left| \hf_\eta' \right|\left(\frac{\xib_{\eps}}{\xi_{*}}\right)\right)(\zeta^{\eps,n},x^{\eps,n})
                            + \frac{\eps^2c_f}n.
            $$
            Recalling from Lemma \ref{lem: construction of h eta} that $|x||\hf'_\eta(x)|\le\eta$ for $x\in \R$, we finally obtain
            \beq\label{eq: supersol Ac gradient constraint not satisfied}\bal
                \left| 1+\psi^{\eps,\eta,n}_x {\left(\zeta^{\eps,n},x^{\eps,n}\right)}\right|&\le
                    \eps^3(1-\eta^2)+\frac{\eps^2c_f}n{<} \eps^3\qquad \pourtout n{\ge 1+ \frac{c_f}{\eps\eta^2}=:N_{\eps,\eta}}.
            \eal\eeq

        \step5{We now   show that  $\{\xib_{\eps}(\zeta^{\eps,n},x^{\eps,n})\;;\; \eps\in(0,\eps_\eta],n\ge N_{\eps,\eta}\}$ is uniformly bounded.}

            We first appeal to Lemma \ref{lem: remainder estimate}, {recall Assumption \ref{ass: pi smooth},}  Lemma \ref{lem: existence and property varpi} and that $(\zeta^{\eps,n},$ $n\ge 1,$ $\eps\in(0,\eps_{\eta}])$ is bounded, see step 3. Since   $\phi^\eps$ does not depend on the $x$-variable, this implies
            \b*
               -\eps^{-2}(\LSX + \LPlSXh)\psi^{\eps,\eta,n}
                &=&
                    -\frac12\frac{\pi_{pp}}{\left(\pi_p\right)^2}\sigma^2\xib_{\eps}^{2}
                   -\Hc\phi^\eps - \frac{1-\eta}2\sigma^2\delta^2(\varpi H)_{\xi\xi}\circ\xib_{\eps}  +R^{\eps ,n}\;,
            \e*
            at the point $(\zeta^{\eps,n},x^{\eps,n})$, in which,{ by (ii) of Lemma \ref{lem: remainder estimate},}
            \beq\label{eq: supersol Ac remainder}
                |R^{\eps,n}| \le C_{\eta}(1+\eps|\xib_{\eps}|+\eps^2|\xib_{\eps}|^2)(\zeta^{\eps,n},x^{\eps,n}){,}
            \eeq
            for some $C_{\eta}>0$ independent on $n$ and $\eps$.
            By \reff{eq: supersol Ac spersol for psi eps eta n}, we then have
            \beq\label{eq: supersol Ac inequality for xi bounded}
                \left(\frac12\frac{\pi_{pp}}{\left(\pi_p\right)^2}\sigma^2 \xib_\eps^2\right)(\zeta^{\eps,n},x^{\eps,n})-\left|R^{\eps,n}\right|
                    \le-\left(\Hc\phi^\eps + \frac{1-\eta}2\sigma^2\delta^2(\varpi H)_{\xi\xi}\circ \xib_{\eps}\right)(\zeta^{\eps,n},x^{\eps,n}).
            \eeq
            We first consider the last term of the previous inequality.
            By Lemma \ref{lem: existence and property varpi} and the boundedness of $(\zeta^{\eps,n},\eps\in (0,\eps_{\eta}],n\ge 1)$,  we can find {$C_{\eta}>0$}, independent on $n$, $\eps$ and $\eta$,  such that
            such that $|\varpi_{\xi\xi}\circ\xib_{\eps}|(\zeta^{\eps,n},x^{\eps,n})\le  C$. The same Lemma and Remark \ref{rem: property varpi}
            also imply  that $|\varpi_\xi\circ\xib_{\eps}|(\zeta^{\eps,n},x^{\eps,n})\le1$ and $|\varpi\circ\xib_{\eps}|(\zeta^{\eps,n},x^{\eps,n})\le |\xib_{\eps}(\zeta^{\eps,n},x^{\eps,n})|$. Using Lemma \ref{lem: construction of h eta}, and the fact that $\xi_{*}\ge 1$ and $\eta\le 1$, it follows that, at the point $(\zeta^{\eps,n},x^{\eps,n})$,
            $$\bal
                \left|(\varpi H_{\eta})_{\xi\xi}\right| \circ\xib_{\eps}
                &= \left| \varpi_{\xi\xi}H_{\eta} + 2\varpi_\xi H_\eta' + \varpi H^{''}_{\eta} \right|\circ\xib_{\eps}\\
                &\le   {C_{\eta}}
                    +\frac2{\xi_{*}}\left| \hf_\eta'\left(\frac{\xib_{\eps}}{\xi_{*}}\right) \right|
                        \1_{[\xi_{*},c_\eta\xi_{*}]}(|\xib_{\eps}|)
                    + \frac{\left|\xib_{\eps}\right|}{(\xi_{*})^{2}} \left| \hf_\eta''\left(\frac{\xib_{\eps}}{\xi_{*}}\right) \right|\\
                &\le    {C_{\eta}}
                    + \frac{2|\xib_{\eps}|}{(\xi_{*})^{2}}\left| \hf_\eta'\left(\frac{\xib_{\eps}}{\xi_{*}}\right) \right|
                    + \frac{\left|\xib_{\eps}\right|}{(\xi_{*})^{2}} \left| \hf_\eta''\left(\frac{\xib_{\eps}}{\xi_{*}}\right) \right| \\
                &\le    {C_{\eta}}
                    + \frac{2}{\xi_{*}}(\eta   + C^*)\\
                &\le  {C_{\eta}}+2(1  + C^*)=: {\bar C_{\eta}}.
            \eal$$
            Plugging this result into \reff{eq: supersol Ac inequality for xi bounded} leads to
             \b*
                 \left(\frac12\frac{\pi_{pp}}{\left(\pi_p\right)^2}\sigma^2\xib_\eps^2\right)(\zeta^{\eps,n},x^{\eps,n})-\left|R^{\eps,n}\right|
                  \le-\left(\Hc\phi^\eps {-} \bar C\frac{1-\eta}2\sigma^2\delta^2\right)(\zeta^{\eps,n},x^{\eps,n}).
            \e*
            The later combined with   Assumption \ref{ass: pi smooth}, \reff{eq: hyp mu sigma}, \reff{eq: supersol Ac remainder} and the fact that both $\zeta^{\eps,n}$
            and $\zeta^\eps$
            lie in $B_{r_o}(\zeta_o)$, {and the identity  $\eps \xib_{\eps}(\zeta^{\eps,n},x^{\eps,n})= x^{\eps,n}-\theta(\zeta^{\eps,n})$}, allows us  to  find  a constants $K_{\eta}>0$, independent on $n$ and $\eps$,  such that
            $$
               \left[ \left(\xib_{\eps}\right)^2
                    - K_{\eta}\left( 1+\left|\eps\xib_\eps\right| + \left|\eps\xib_\eps\right|^2 \right) \right](\zeta^{\eps,n},x^{\eps,n})\le0.
            $$
            This proves our claim.

        \step6{We are now in position to conclude the proof{.}}

            By the previous step,
            for all $\eps\in(0,\eps_\eta]$, we may assume, after possibly passing to a subsequence, that              $(\zeta^{\eps,n},x^{\eps,n},\xib_{\eps}(\zeta^{\eps,n},x^{\eps,n}))\rightarrow
            (\bar \zeta^\eps,\theta(\bar \zeta^{\eps}),\bar \xi^\eps)\in {\D}\x \R^{2}$ as $n\rightarrow\infty$.
            Classical arguments then show that   $(\bar \zeta^\eps,\bar \xi^\eps)\rightarrow(\zeta_o,\hat\xi)$ for some bounded $\hat\xi\in \R$, and therefore $\theta(\bar \zeta^{\eps})\to \theta(\zeta_{o})=x_{o}$,  as $\eps\to 0$, after possibly passing to a subsequence.
	Moreover,  (i) of Lemma \ref{lem: remainder estimate} now implies that $R^{\eps,n}\to 0$ as $n\to \infty$ and then $\eps\to 0$.  Hence, sending $n\rightarrow\infty$ and then $\eps\rightarrow0$ in \reff{eq: supersol Ac inequality for xi bounded}
          provides
            $$
                \frac12\left(\frac{\pi_{pp}}{\left(\pi_p\right)^2}\sigma^2\right)(\zeta_o)\hat \xi^2
                    \le
                -\Hc\vp(\zeta_o) - \frac{1-\eta}2\{\sigma^2\delta^2(\varpi H_{\eta})_{\xi\xi}\}(\zeta_o,\hat\xi).
            $$
           The same arguments as in step 5 then shows that
            $$
                \hat\xi^2
                    \le
                    \left(\frac{
                -\Hc\vp + \sigma^2\delta^2\bar C(1-\eta)/2}{\frac12\frac{\pi_{pp}}{\left(\pi_p\right)^2}\sigma^2}
                    \right)(\zeta_o).
            $$
            We now choose $\xi_{*}\ge 1$ defined by
            \be\label{eq: def xi star}
            (\xi_{*})^{2}:= 2\vee 2\left(\frac{
                -\Hc\vp + \sigma^2\delta^2\bar C/2}{\frac12\frac{\pi_{pp}}{\left(\pi_p\right)^2}\sigma^2}.
                    \right)(\zeta_o).
            \ee
            Note that all the quantities on the right-hand side are given a-priori.
            Then,
            {
            $|\hat \xi|< \xi_{*}.$
            }
            In particular,  $H_{\eta}=1$  in a neighborhood of $\hat \xi$, see Lemma \ref{lem: construction of h eta}, and the above then implies that
            $$
                    \frac12\left(\frac{\pi_{pp}}{\left(\pi_p\right)^2}\sigma^2\right)(\zeta_o)\hat\xi^2
                    \le-\Hc\vp(\zeta_o)
                        - \frac{1-\eta}2\left(\sigma^2\delta^2\varpi_{\xi\xi}\right)\left(\zeta_o,\hat\xi\right).
            $$
            Since $\varpi$ is solution of \reff{eq: first corrector equation}, it follows that
            $$
                \Hc\vp(\zeta_o)\le -h(\zeta_o) + \frac\eta2\sigma^2\delta^2\varpi_{\xi\xi}(\zeta_o,\hat\xi).
            $$
           {  It remains to let $\eta\rightarrow0$ and recall from Lemma \ref{lem: existence and property varpi} that $|\varpi_{\xi\xi}(\zeta_o,\cdot)|$ is bounded.}

        \qed

    \subsection{The Terminal condition}\label{sec: terminal condition}

   \begin{proposition}\label{prop: terminal condition for u}  Let the conditions of Theorem \ref{thm: main} hold. Then, $u^\ast=u_\ast=0$ on $\DT$.
        \end{proposition}

        \proof The fact that $u_{*}(T,\cdot)\ge 0$ follows from Remark \ref{rem: v le veps}.
            In the following, we prove that $u^{*}(T,\cdot)\le 0$.
            We assume to the contrary that we can find ${(T,s_{o},p_{o})}:=\zeta_{o}\in{\DT} $ such that
            \beq\label{eq: terminal sub sol contradiction}
                u^\ast(\zeta_o)\ge4{\kappa}
                \quad\mbox{for some}\quad{\kappa}>0
            \eeq
            and work towards a contradiction.

            \step1{We construct a test function $\psi^{\eps}$ for $v^{\eps\ast}$ and show that $v^{\eps\ast}-\psi^\eps$ admits a local maximizer
            $(\tilde t_{\eps},\tilde s_{\eps},\tilde p_{\eps},\tilde x_{\eps})=(\tilde\zeta_\eps,\tilde x_\eps)\in \Di\x \R$.
            }

          	By Lemma \ref{lem: u does not depend on x},
          there are $(\zeta_\eps)_{\eps>0}\subset \D$ and $x_o\in\R$ such that
            \beq\label{eq: terminal convergence points}
                 \zeta_\eps\underset{\eps\downarrow0}{\longrightarrow}\zeta_o,
                \quad x_\eps:= \theta(\zeta_\eps)\underset{\eps\downarrow0}{\longrightarrow}\theta(\zeta_{o})=:x_o
                \quad \mbox{and}\quad
                  u^{\eps\ast}(\zeta_\eps,x_\eps)\underset{\eps\downarrow0}{\longrightarrow}  u^*(\zeta_o)\;,
            \eeq
            in which $(t_{\eps},s_{\eps},p_{\eps}):=\zeta_{\eps}$.
            Note that, after possibly passing to a subsequence, one can assume that
            \be\label{eq: proof condition T teps<T}
            \zeta_{\eps}\in \Di\; \mbox{ for all }\; \eps>0.
            \ee
            Indeed, Theorem  \ref{thm: pde veps} and Theorem \ref{thm: pde v}  imply that
            $$
                u^{\eps*}(\zeta,x)  \le \eps |x|\; \mbox{ for all }\; (\zeta,x)\in {\DT}\x \R,
            $$
            which would lead to a contradiction of \reff{eq: terminal sub sol contradiction} if \reff{eq: proof condition T teps<T} was not satisfied,
            at least along a  subsequence, since, by \reff{eq: terminal convergence points},
            $(\zeta_\eps,x_\eps)_{\eps>0}$ is   bounded.

            Combining arguments similar to those of the proof of Proposition  
            \ref{prop : subsol u*} (Step 1) with
            \reff{eq: terminal sub sol contradiction},
            \reff{eq: terminal convergence points},
            Assumptions \ref{ass: u eps finite} and \ref{ass: pi smooth}   allow us to construct 
            $0<r_o\le\bar r_o, \eps_o\in(0,1], c_o>0$ and $\iota>0$ such that, for all $\eps\in(0,\eps_o]$,
            \be
                &\displaystyle
                    (\zeta_\eps,x_\eps)\in B_{o,\frac12}
                \quad\mbox{and}\quad
                 u^{\eps\ast}(\zeta_\eps,x_\eps)\ge2{\kappa},\label{eq: terminal localisation of seq}&\\
                &\displaystyle
                \pi_p\ge2\iota \quad\mbox{on}\quad B_o,\label{eq: terminal control of pi_p}&\\
                &\displaystyle
                    u^{\eps\ast}-\bar\phi(\cdot;s_\eps,p_\eps)<0\quad\mbox{on}\quad B_o\backslash B_{o,\frac12}
                ,\label{eq: terminal diff u eps and bar phi}&
            \ee
            where
            $B_{o}:=[T-r_o,T]\x \bar B_{r_{o}}(s_o,p_o)\x\bar B_{\bar r_o}(x_o)$,
            \b*
                &\displaystyle B_{o,\frac12}:=
                    \left\{
                        (\zeta,x)\in B_o
                            \st \zeta\in[T-\frac{r_o}2,T]\x
                                \bar B_{\frac{r_o}2}(s_o,p_o)
                                \mbox{ and }x\in\bar B_{\frac{\bar r_o}2}(x_o)
                    \right\},&\\
                &\displaystyle {\bar\phi(\cdot;s_\eps,p_\eps):(t,s,p,x)\in\D\x\R\longmapsto
                    c_o\left(\abs{s_{\eps}-s}^4+\abs{p_{\eps}-p}^4+\abs{x-\theta(t,s,p)}^2\right).}&
            \e*
            Recalling \reff{eq: proof condition T teps<T} and Assumption \ref{ass: pi smooth},
            we may then define, for each $\eps\in(0,\eps_o]$, the smooth function $\psi^\eps:= v+\eps^2\phi^\eps$ with
            $$
                \phi^\eps:(t,s,p,x)\in {\D}\x\R\longmapsto {\kappa}\frac{T-t}{T-t_\eps}+ \bar \phi(t,s,p,x;s_\eps,p_\eps).
            $$
            By the upper semi-continuity of $v^{\eps\ast}$,
            we deduce from \reff{eq: terminal localisation of seq}
            and \reff{eq: terminal diff u eps and bar phi} that $v^{\eps\ast}-\psi^\eps$ admits on $B_o$ a local maximizer
            $(\tilde\zeta_\eps,\tilde x_\eps)\in B_{o,\frac12}$ for every $\eps\in(0,\eps_o]$, and that moreover
            $$
                  u^{\eps\ast}(\tilde\zeta_\eps,\tilde x_\eps)\ge{\kappa} .
            $$
           By the argument used above, this implies that $\tilde\zeta_\eps\in \Di$ for all $\eps\in(0,\eps_o]$ after possibly choosing a subsequence.

            \step2{We now show that $(\xib_\eps(\tilde\zeta_\eps,\tilde x_{\eps}))_{\eps\in(0, \eps_o]}$ is uniformly bounded.}

             We fix $\eps\in(0,\eps_o]$.   The previous step and Theorem \ref{thm: pde veps} imply that
                \beq\label{eq: terminal pde for psi eps}
                \displaystyle \max\left\{
                    -(\LSX+\LPlSXh)\psi^{\eps}
                    \;;\;
                    -\eps^3+1+\psi^\eps_x
                    \;;\;
                    -\eps^3-1-\psi^\eps_x
                \right\}(\tilde\zeta_\eps,\tilde x_\eps)
                \le0.
            \eeq
                Straightforward computations based on the gradient constraints give
                \beq\label{eq: terminal wi bounded}
                    -\frac1{2c_o}\le \xib_\eps(\tilde \zeta_{\eps},\tilde x_{\eps})\le\frac1{2c_o}.
                \eeq

                \step3{We can now conclude the proof.}

                 We fix $\eps \in (0,\eps_{o}]$ and   focus on the second order operator in \reff{eq: terminal pde for psi eps}.
                It follows from \reff{eq: terminal control of pi_p} that
                $
                    \psi^\eps_p(\tilde\zeta_\eps,\tilde x_\eps)\ge \iota>0,
                $
                after possibly changing $\eps_{o}$.
                Hence,  {Step 2 and  (i) of }Lemma \ref{lem: remainder estimate} imply that
                $$
                    \left(-\frac12\frac{ \pi_{pp}}{(\pi_p)^2}\sigma^2\xib_\eps^2-\Hc\phi^\eps-{\LXlSP^{\hat a}}\phi^\eps+R^\eps\right)
                        (\tilde \zeta_\eps,\tilde x_\eps)\le 0\;,
                $$
                where $\sup_{\eps\in (0,\eps_{o}]}   \abs{R^\eps}(\tilde\zeta_\eps,\tilde x_\eps)<\infty$.
                Recalling  {\reff{eq: terminal wi bounded}}, the fact that {$(\tilde \zeta_{\eps}$, $\eps\in (0,\eps_{o}])$} is bounded, { that $\tilde x^{\eps}=\eps \xib_{\eps}(\zeta_{\eps})+\theta(\zeta_{\eps})$}, {and Assumption \ref{ass: pi smooth}}, we finally deduce that
                $$
                    \frac{{\kappa}}{T-t_\eps}\le
                        \left(
                            \frac12\frac{ \pi_{pp}}{(\pi_p)^2}\sigma^2\frac1{4c_o^2}+
                                \left(\Hc+\LXlSP^{\hat a}\right)\bar\phi(\cdot;s_\eps,p_\eps)+R^{\eps}
                        \right)\left(\tilde\zeta_\eps,\tilde x_\eps\right)
                        \le C\pourtout\eps\in(0,\bar\eps_o]\;,
                $$
                for some constant $C>0$ (independent of $\eps$). As $ t_{\eps}\to T$, we obtain a contradiction.
                \qed


\section{Explicit resolution of the first corrector equation} \label{sec: proof lem existence and property varpi}

In this section, we prove Lemma \ref{lem: existence and property varpi}.   We follow the steps of \cite{SoTo12}. Namely, we look for a solution of the first order equation \reff{eq: first corrector equation} with an additional condition at the boundary $\xi=0$. We fix $\zeta \in \D$ and simply write $\varpi(\xi)$ for $\varpi(\zeta,\xi)$.   We recall that we work under  Assumption \ref{ass: pi smooth}.

 It is natural to search for a solution of the form
    $$
        \varpi(\xi)=
        \left\{
            \begin{tabular}{ll}
              $k_4\xi^4 + k_2\xi^2+k_1\xi$ & $\xi_1\le\xi\le\xi_0$\;,\\
              $-\xi + k_3$ & $\xi\le\xi_1$\;,\\
              $\xi + k_0$ & $\xi\ge\xi_0\;,$
            \end{tabular}
        \right.
    $$
   for some real numbers $k_{4}, k_{3},k_{2},k_{1},k_{0}$ and $\xi_{1}\le \xi_{0}$.
    Since the fourth order polynomial solves the second order equation, we find
    \beq\label{eq: solution second order}
        k_4 = -\frac1{12}\frac{\pi_{pp}}{\delta^2\left(\pi_p\right)^2}
        \qquad \mbox{and} \qquad
        k_2 = \frac{h}{\sigma^2\delta^2}.
    \eeq
    If we now assume that $\varpi_{\xi\xi}$ is continuous at the point $\xi_0$ and $\xi_1$, we have
    $$
        12k_4(\xi_0)^2 +2k_2 = 12k_4(\xi_1)^2 +2k_2 = 0,
    $$
    that is
    $$
        (\xi_0)^2=(\xi_1)^2=
            {2\frac{h}{\sigma^2}}\x
            {\frac{\left(\pi_p\right)^2}{\pi_{pp}}}\;,
    $$
    which, by the fact that $\pi_{pp}>0$, implies that $h\ge0$ and
    $$
        \hat \xib:=\xi_0=-\xi_1=\left(
            {2\frac{h}{\sigma^2}}\x
            {\frac{\left(\pi_p\right)^2}{\pi_{pp}}} \right)^{\frac12}.
    $$
    Assuming now that $\varpi_{\xi}$ is continuous at the point $\xi_0$ and $\xi_1$ leads to
    \beq\label{eq: smooth first derivative}\begin{matrix}
        \displaystyle4k_4( \hat \xib)^3 +2k_2  \hat \xib+k_1 = 1\;,\\
        \displaystyle-4k_4( \hat \xib)^3 -2k_2  \hat \xib+k_1 = -1\;,
    \end{matrix}\eeq
    which gives $k_1=0$.
    By substituting \reff{eq: solution second order} into \reff{eq: smooth first derivative},
    $$
        \displaystyle-\frac{\pi_{pp}}{\delta^2\left(\pi_p\right)^2}( \hat \xib)^3 +\frac{6h}{\sigma^2\delta^2}  \hat \xib = 3.
    $$
    Since, by the above,
    \be\label{eq: explicit form of h}
        h=\frac{\sigma^2\pi_{pp}}{2(\pi_p)^2}( \hat \xib)^2,
    \ee
    we obtain
    \be\label{eq: explicit form of xib}
         \hat \xib = \left(\frac32\frac{\delta^2\left(\pi_p\right)^2}{\pi_{pp}}\right)^{\frac13}.
    \ee
    The remaining constants $k_0$ and $k_3$ are obtained by assuming the continuity of $\varpi$ at the points $\xi_0$ and $\xi_1$. Gathering the above terms together, we finally obtain
    \beq\label{eq: explicit form of varpi}
        \varpi(\xi)=
        \left\{
            \begin{tabular}{ll}
              $-\frac1{8\hat \xibu^3}\xi^4 + \frac{3}{4\hat \xibu}\xi^2$ & $-\hat \xib\le\xi\le\hat \xib$\;,\\
              $-\xi - \frac{3\hat \xibu}8$ & $\xi\le-\hat \xib$\;,\\
              $\xi - \frac{3\hat \xibu}8$ & $\xi\ge\hat \xib\;.$
            \end{tabular}
        \right.
    \eeq
    The remaining properties stated in Lemma \ref{lem: existence and property varpi}  are straightforward under Assumption \ref{ass: pi smooth}.


\section{Verification of the  assumptions in the examples }\label{sec: proof examples}

 In this section, we provide the proofs of Propositions  \ref{prop: ass ok for expo BS}    and \ref{prop: ass ok for powe BS}. We  also explain how to construct an explicit {\sl almost} optimal strategy.

\subsection{Exponential case }

 We provide here the proof  of Proposition \ref{prop: ass ok for expo BS}.

 {\bf Proof of Proposition \ref{prop: ass ok for expo BS}}  First note that \reff{eq: pi explicit in BS example} together with   Assumption \ref{ass: BS expo case} imply Assumption \ref{ass: pi smooth}.
   Under the boundedness condition   b.~of Assumption \ref{ass: BS expo case}, the function $h$ is bounded, see \reff{eq: expo BS case - explicit functions}. It follows that the map defined in \reff{eq: def bar u example} is bounded. Moreover, standard arguments show that comparison holds in the viscosity solution sense for the above equation in the class of   functions with polynomial growth, see \cite{CrIsLi92}.  Then, Assumption \ref{ass: comparison for u} will hold if one shows that  there exists $C>0$ such that
   \be\label{eq: uniforml x linear growth ueps}
    0\le u^{\eps}(\zeta,x)\le C(1+\eps|x|) \;\mbox{ for all } (\zeta,x)\in \D\x \R \;\mbox{ and } \;\eps\in (0,1],
   \ee
   in which the left-hand side inequality is already a consequence of Remark \ref{rem: v le veps}.
  This will also imply Assumption  \ref{ass: u eps finite}.
 The following arguments aim at proving the right-hand side inequality of  \reff{eq: uniforml x linear growth ueps}.

{\sl Step 1. }  We restrict to $0<\eps\le 1$. Set
       \be\label{eq: def psi eps proof expo}
       \psi^{\eps}(t,s,p,x):=v(t,s,p,x)+\eps^{4}\check\varpi\circ \xib_{\eps}(t,s,x) \;\mbox{ for } (t,s,p,x)\in \D\x \R,
       \ee
in which $\check \varpi$ is the solution of {\reff{eq: first corrector equation} as constructed in Section \ref{sec: proof lem existence and property varpi}} but for $\delta=\sigma=1$ and $\pi_{p}^{2}/\pi_{pp}=1$. For later use, observe that it takes non-negative values. We denote by $\check \xib$ the corresponding $\hat \xib$ and $\check h$ the corresponding $h$. Then, $\check \xib$ and $\check h$ are constant, and $\check \varpi$ depends only on $\xi$.
Let us also define
       \b*
       \hat a^{\eps}&:=&\frac{-\bar \sigma_{0}^{\top} D\psi^\eps}{\pi_p}
           \\
           & =&
          {\eta}p\sigma\left[(\theta-x)(1-\eps^{3}\varpi_{\xi}\circ\xib_{\eps}) +\eps^{3} \varpi_{\xi}\circ\xib_{\eps} (\frac{\lambda}{\sigma \eta}- s^{2} \bar \pi_{ss} ) -\frac{\lambda}{\sigma \eta}  \right]
       \e*
        and
                    \be
                    \Jc_{\eps}&:=&\{(t,s,x)\in[0,T]\x (0,\infty)\x \R: -\check \xib(t,s) <\xib_{\eps}(t,s,x)<\check \xib(t,s)\}\nonumber\\
                    &=&\{(t,s,x)\in[0,T]\x (0,\infty)\x \R: -\eps \;\check \xib(t,s)< x-\theta(t,s)<\eps \;\check \xib(t,s)\},\label{eq: def domain reflexion Skorohod problem}
                    \ee
        recall Proposition \ref{prop: pi explicit BS example} and \reff{eq: do not depend on x and p}.
        Lemma \ref{lem: existence and property varpi}
        allows one to characterize the boundaries of this domain in terms of the function {$\varpi$}:
        \be
            \partial\Jc^\pm_{\eps}:=
            \left\{
                \xib_{\eps}=\mp\check \xib
            \right\}\subset\{\varpi_{\xi}\circ \xib_{\eps}=\mp 1\}.\label{eq: def boundary reflexion Skorohod problem}
        \ee
For later use, note that Assumption \ref{ass: BS expo case} implies that
	\be\label{eq: x bounded in Jceps and hat a CL lipsch}
	(t,s,x)\in \Jc_{\eps}\Longrightarrow \left\{ |x|\le C_{K} \;\mbox{ and }\; |p^{-1}\hat a^{\eps}(t,s,x,p)|\le C_{K} \mbox{ for all $p<0$}\right\},
	\ee
	in which $C_{K}$ denotes from now on a generic positive constant which depends only on the constant $K>0$ of Assumption \ref{ass: BS expo case}, and that may change from line to line.

        We now fix $(t_{o},s_{o},x_{o})$ in the closure of {$\Jc_{\eps}$}. The general case will be discussed in the last step of the proof. We define     $(X^{\eps},L^{\eps})$ as the solution of the following Skorokhod problem
        \beq\label{eq: skorokhod problem} \left\{\bal
                &X^{\eps}
                    = x_{o}+ \int_{t_{o}}^\cdot X^{\eps}_{\tau}\frac{dS_{\tau}}{S_{\tau}} + \int_{t_{o}}^\cdot dL^{\eps+}_\tau-\int_{t_{o}}^\cdot dL^{\eps-}_{\tau}\;,\\
                &(\cdot,S,X^{\eps})\in\Jc_\eps\quad dt\otimes d\P\mbox{-a.e. on }[t_{o},T]\;,\\
                &L^{\eps\pm}=\int_{t_{o}}^\cdot\chi_{\{(\tau,S_\tau,X^{\eps}_\tau)\in\partial\Jc_{\eps}^\pm\}}dL^{\eps\pm}_\tau\;,
        \eal\right.\eeq
        in which $S=S^{t_{o},s_{o}}$ and $L^{\eps}=L^{\eps+}-L^{\eps-}$ where  $L^{\eps+},L^{\eps-}$ are continuous and non-decreasing.
To see that the above admits a solution, first observe that Assumption \ref{ass: BS expo case} ensures that we can find $\kappa\in \R$ such that $-\check \xib+\theta>\kappa$ on $[0,T]\x (0,\infty)$. Hence, the process $X^{\eps}$ satisfies the above if and only if $X^{\eps}-\kappa>0$, in which case
$$
X^{\eps}-\kappa=(x_{o}-\kappa)\exp\left(\int_{t_{0}}^{\cdot}(\mu -\textstyle{\frac12}\sigma^{2})d\tau+\Int_{t_{0}}^{\cdot}d W_{\tau}+\Int_{t_{0}}^{\cdot}d\bar L^{\eps+}_{\tau}-\int_{t_{0}}^{\cdot}d\bar L^{\eps-}_{\tau}\right)\;\;\;\mbox{on } [t_{o},T]\;,
$$
with  $d\bar L^{\eps\pm}=dL^{\eps \pm}/(X^{\eps}_{\tau}-\kappa)$. Thus, solving    \reff{eq: skorokhod problem} is equivalent to finding the solution $(\bar X^{\eps},\bar L^{\eps})$ of the Skorohod problem
  \beq\nonumber
  \left\{\bal
                &\bar X^{\eps}
                    = \ln(x_{o}-\kappa)+ \int_{t_{0}}^{\cdot}(\mu -\textstyle{\frac12}\sigma^{2})d\tau+\Int_{t_{0}}^{\cdot}d W_{\tau}+\Int_{t_{0}}^{\cdot}d\bar L^{\eps+}_{\tau}-\Int_{t_{0}}^{\cdot}d\bar L^{\eps-}_{\tau}\;, \\
                &   U^{-}\le \bar X^{\eps}\le U^{+} \quad dt\otimes d\P\mbox{-a.e. on }[t_{o},T]\;,\\
                &\bar L^{\eps\pm}=\int_{t_{o}}^\cdot\chi_{\{\bar X^{\eps}_\tau=U^{\pm}\}}dL^{\eps\pm}_\tau\;,
        \eal\right.\eeq
        in which
        $$
       U^{\pm}:= { \ln}\left(-\kappa + (\pm \eps \;\check \xib+\theta)(\cdot,S)\right).
        $$
 Existence now follows from \cite[Lemma 6.14]{KaSc88}, see the constructive proof for the fact that the solution is adapted.

 We next define $(Y^{\eps},P^{\eps})$ as the solution of
        \begin{equation}
         Y^{\eps}
                    = y_{o}- \int_{t_{o}}^\cdot (1+\eps^{3})dL^{\eps+}_\tau + \int_t^\cdot (1-\eps^{3})dL^{\eps-}_\tau
            \;,\;
                P^{\eps}=p_{o}+\int_{t_{o}}^\cdot \hat a^\eps\left(\tau,S_\tau,P^{\eps}_\tau,X^{\eps}_\tau \right)dW_\tau\;,
                \label{eq: example definition P eps}
        \end{equation}
       in which $p_{o}<0$ and     $y_{o}:=\psi^{\eps}(t_{o},s_{o},p_{o},x_{o})+c$ for some $c>0$ to be chosen later on.
  The existence of a unique strong solution to \reff{eq: example definition P eps} follows from \reff{eq: x bounded in Jceps and hat a CL lipsch}, the process $P^{\eps}$ is a martingale.

 {\sl Step 2.}       We now apply It\^{o}'s Lemma to $\psi^{\eps}$. The definition of $\hat a^{\eps}$ and  the above dynamics lead to
        $$
        \bal
        Y^{\eps}_{T}-\psi^{\eps}(T,S_{T},P^{\eps}_{T},X^{\eps}_{T})
          &\quad =c
                    -\int_{t_{o}}^T\left(\LSX+\LPlSXh\right)\psi^\eps\left(\tau,S_\tau,P^{\eps}_\tau ,X^{\eps}_\tau\right)d\tau\\
                    &\quad\quad-\int_{t_{o}}^T\left[(1+\eps^{3})+\psi^\eps_x\left(\tau,S_\tau,P^{\eps}_\tau,X^{\eps}_\tau \right)\right]
                        dL^{\eps+}_\tau\\
                    &\quad\quad+\int_{t_{o}}^T\left[(1-\eps^{3})+\psi^\eps_x\left(\tau,S_\tau,P^{\eps}_\tau,X^{\eps}_\tau \right)\right]
                        dL^{\eps-}_\tau\\
                &\quad\ge c-\int_{t_{o}}^T\left(\LSX+\LPlSXh\right)\psi^\eps\left(\tau,S_\tau,P^{\eps}_\tau,X^{\eps}_\tau \right)d\tau\\
                    &\quad\quad-\eps^3\int_{t_{o}}^T\left[1+\check\varpi_\xi\circ\xib_\eps\left(\tau,S_\tau,X^{\eps}_\tau \right)
                     \right]
                        dL^{\eps+}_\tau\\
                    &\quad\quad+\eps^3\int_{t_{o}}^T\left[-1+\check\varpi_\xi\circ\xib_\eps\left(\tau,S_\tau,X^{\eps}_\tau \right)
                     \right]
                        dL^{\eps-}_\tau.
            \eal
            $$
         We next appeal to    \reff{eq: def boundary reflexion Skorohod problem} and  the characterization of $L^{\eps+},L^{\eps-}$ in \reff{eq: skorokhod problem}  to provide a lower bound to the last expression:
        \be\label{eq: YT-psieps T}
        Y^{\eps}_{T}-\psi^{\eps}(T,S_{T},P^{\eps}_{T},X^{\eps}_{T})
        \ge c-\int_{t_{o}}^T\left(\LSX+\LPlSXh\right)\psi^\eps\left(\tau,S_\tau,P^{\eps}_\tau,X^{\eps}_\tau \right)d\tau=:c-\eps^{2} E_{\eps}.
         \ee
         We first consider the left-hand side term.
        The definition of $\psi^{\eps}$ and the identities   $v(T,s,p,x)=g(s)-x-\frac1\eta \ln(-p)$,
        see Proposition \ref{prop: pi explicit BS example},     lead to
        \b*
        Y^{\eps}_{T}+\ell^{\eps}(X^{\eps}_{T})-g(S_{T})+\frac1\eta \ln(-P^{\eps}_{T})&\ge&
        Y^{\eps}_{T}-\psi^{\eps}(T,S_{T},P^{\eps}_{T},X^{\eps}_{T}) +\psi^{\eps}(T,S_{T},P^{\eps}_{T},X^{\eps}_{T})\\
        &&+\ell^{\eps}(X^{\eps}_{T})-g(S_{T})+\frac1\eta \ln(-P^{\eps}_{T})\\
        &\ge&
        Y^{\eps}_{T}-\psi^{\eps}(T,S_{T},P^{\eps}_{T},X^{\eps}_{T})+\eps^{4}\check\varpi\circ \xib_{\eps}(T,S_{T},X^{\eps}_{T})\\
        & &-\eps^{3}|X^{\eps}_{T}| .
        \e*
     Recall that $\check\varpi\ge 0$. {We also know from \reff{eq: x bounded in Jceps and hat a CL lipsch} and \reff{eq: skorokhod problem}  that $|X^{\eps}_{T}|\le C_{K}$. }Hence, we deduce from the  above that
        \be
        Y^{\eps}_{T}+\ell^{\eps}(X^{\eps}_{T})-g(S_{T}) +\frac1\eta \ln(-P^{\eps}_{T})    &\ge&
        Y^{\eps}_{T}-\psi^{\eps}(T,S_{T},P^{\eps}_{T},X^{\eps}_{T}) - C_{K}\eps^{3} . \label{eq: decompo 1 - minoration of final gain}
        \ee
     We now consider the right-hand side term in \reff{eq: YT-psieps T}.
   Since $\pi_{p}>0$ and $\check \varpi$ do not depend on $p$, one can apply the expansion of Lemma \ref{lem: remainder estimate}.
     It implies
     \begin{equation}\label{eq: def Eeps proof expo}
             E_{\eps}=
            \int_{t_{o}}^T \left(  \frac{\sigma^{2}}{2}\eta \xib_{\eps}(\tau,S_{\tau},X^{\eps}_{\tau})^{2} + \frac{\sigma^{2}}{2} \delta^2(\check \varpi_{\xi\xi}\circ\xib_{\eps})(\tau,S_{\tau},X^{\eps}_{\tau})+R_{\eps}(\tau,S_{\tau},P^{\eps}_{\tau},X^{\eps}_{\tau})  \right) d\tau
            \end{equation}
     in which the map $R_{\eps}$ is given by \reff{eq : def R eps iota} for $\phi:=0$ and $w:=\check\varpi$.

            Direct computations based on  condition b.~of Assumption \ref{ass: BS expo case}, the specific forms of $\hat a$ and $\pi$,  and \reff{eq: x bounded in Jceps and hat a CL lipsch} lead to $|R_{\eps}|\le C_{K}$ on the closure of  $\Jc_{\eps}$, and therefore:
            $
            |R_{\eps}(\cdot,S,P^{\eps},X^{\eps})| \le C_{K}.
            $
            It also follows from Assumption \ref{ass: BS expo case}, \reff{eq: expo BS case - explicit functions} and \reff{eq: def domain reflexion Skorohod problem} that $|\xib_{\eps}(\cdot,S,X^{\eps})|\le C_{K}$. Finally
            \reff{eq: explicit form of varpi} {above for the coefficients entering in the definition of  $\check \varpi$  provides} a uniform bound for the remaining term. Therefore
            \be\label{eq: bound Deltaeps in terms of Reps}
            |E_{\eps}|\le C_{K}.
            \ee
Combining \reff{eq: YT-psieps T}, \reff{eq: decompo 1 - minoration of final gain} and \reff{eq: bound Deltaeps in terms of Reps} leads to
	\be\label{eq: Y + ell(X)-g lower bounded in c and P}
	 Y^{\eps}_{T}+\ell^{\eps}(X^{\eps}_{T})-g(S_{T}) \ge c - \frac1\eta \ln(-P^{\eps}_{T})-C_{K}\eps^{2}.
	\ee
     Recall that $C_{K}$ depends only on $K$ but not on $c$. Hence, we can choose $c=(C_{K}+1)\eps^{2}$, and obtain from the previous inequality that
	\b*
	\Psi\left( Y^{\eps}_{T}+\ell^{\eps}(X^{\eps}_{T})-g(S_{T})\right) \ge P^{\eps}_{T}e^{- \eta\eps^{2}}\;,
	\e*
	so that
	\be\label{eq: reach expectation for not admi strat}
	\Esp{\Psi\left( Y^{\eps}_{T}+\ell^{\eps}(X^{\eps}_{T})-g(S_{T})\right)} \ge p_{o}e^{-\eta  \eps^{2}},
	\ee
	since $P^{\eps}$ is a martingale.

	{\sl  Step 3.}
	Note that the strategy $L^{\eps}$ does not  satisfy the admissibility condition \reff{eq: admissibility 
	condition with transaction costs}. However, in Step 4, below we overcome this by replacing  
	 $L^{\eps}$ by an appropriately stopping it (see definition \reff{eq: def Lepsk expo}).
	Towards this goal we start by proving below that the latter inequality implies that
	\be\label{eq: claim proof BS expo case}
	\sup_{L\in \Lmf^{\eps}(t_{o},s_{o},{y_{o},x_{o}})}\Esp{ \Psi(\Delta^{\eps,L})} > p_{o}  ,
	\ee
	in which we abbreviate notations by setting
	$$
	 \Delta^{\eps,L}:=Y^{t_{o},y_{o},\eps,L}_{T}+\ell^{\eps}(X^{t_{o},x_{o},s_{o},L}_{T})-g(S^{t_{o},s_{o}}_{T}).
	$$
	Hence,
	   \b*
	   y_{o}&=&v(t_{o},s_{o},p_{o})
            +\eps^{4}\check\varpi\circ \xib_{\eps}(t_{o},s_{o},{p_{o},x_{o}})+(C_{K}+1)\eps^{2}\\
	   &\ge& v^{\eps}(t_{o},s_{o},{p_{o},x_{o}}),
	   \e*
	   and therefore  	
	\b*
	u^{\eps}(t_{o},s_{o},{p_{o},x_{o}})&=&\eps^{-2}\left(v^{\eps}-v \right)(t_{o},s_{o},{p_{o},x_{o}})\\
	&\le&  \eps^{2}\check\varpi\circ \xib_{\eps}(t_{o},s_{o},{p_{o},x_{o}})+(C_{K}+1).
	\e*
	Recall that Assumption \ref{ass: BS expo case} implies that $\eps{\check \varpi} \circ \xib_{\eps}$ has linear growth in $x$, uniformly in its other variables and in $0<\eps\le 1$, see Remark \ref{rem: property varpi}.  The latter leads to the right-hand side inequality of  \reff{eq: uniforml x linear growth ueps}.
	
	{\sl Step 4.} We now prove our claim \reff{eq: claim proof BS expo case}.
Recalling \reff{eq: x bounded in Jceps and hat a CL lipsch} and the fact that  $g$ is bounded, \reff{eq: Y + ell(X)-g lower bounded in c and P} implies that
$$
Y^{\eps}_{T}+\ell^{\eps}(X^{\eps}_{T})\ge -C_{K}-\int_{t_{o}}^{T} \gamma^{\eps}_{\tau} dW_{\tau}\;,
$$
for some predictable process $\gamma^{\eps}$ which satisfies $|\gamma^{\eps}|\le C_{K}$ for all $0<\eps\le 1$.
Then, it follows from \cite{Kabanov99} that
\b*
Y^{\eps} +\ell^{\eps}(X^{\eps} )&\ge& -C_{K}-\E^{\Q}[\int_{t_{o}}^{T} \gamma^{\eps}_{\tau} dW_{\tau}|\Fc_{\cdot}]\\
&\ge& -C_{K}+ M^{\eps},
\e*
in which $M^{\eps}:=-\int_{t_{o}}^{\cdot} \gamma^{\eps}_{\tau} dW_{\tau}$ satisfies  $\Esp{e^{2\eta \sup_{[t_{o}, T]}|M^{\eps}|}}\le C_{K}$.

 Given $k\ge C_{K}$, we  now denote by $\tau_{k}$ the first time after $t_{o}$ such that $Y^{\eps} +\ell^{\eps}(X^{\eps} )= -k$.   
 Set
 \be\label{eq: def Lepsk expo}
 L^{\eps,k}:=L^{\eps}_{\cdot\wedge \tau_{k}}.
 \ee
 Then, $L^{\eps}$ being continuous, $L^{\eps,k}\in \Lmf^{\eps}(t_{o},s_{o},{y_{o},x_{o}})$ for all $k \ge 1$. Moreover,  since $\Psi\le 0$,
 $$
 \Psi(\Delta^{\eps,L^{\eps}}) -\Psi(\Delta^{\eps,L^{\eps,k}})
 \le  -\Psi(-k) \1_{\{\tau_{k}\le T\}}\le  -\Psi(-k)  \1_{\{\sup_{[t_{o}, T]} |M^{\eps}|\ge k-C_{K}\}}.
 $$
 We next use \reff{eq: reach expectation for not admi strat} and the Markov's inequality to obtain
 $$
 p_{o}e^{-\eta \eps^{2}} \le \Esp{\Psi(\Delta^{\eps,L^{\eps}})}\le \Esp{\Psi(\Delta^{\eps,L^{\eps,k}})}
  -\Psi(-k)  C_{K}/e^{2\eta k}=  \Esp{\Psi(\Delta^{\eps,L^{\eps,k}})}+ C_{K}e^{-\eta k}.
 $$
 Then, taking
 \be\label{eq: def k}
 k:=-\eta^{-1}\ln\left(p_{o}(e^{-\eta \eps^{2}}-1)/C_{K}\right)+1
 \ee
   leads to  \reff{eq: claim proof BS expo case}, recall that $p_{o}<0$.

 {\sl Step 5.} It remains to explain how to consider the general case $(t_{o},s_{o},x_{o})\in [0,T]\x (0,\infty) \x \R$.  First note that an immediate transfer allows one to pass from the initial position $(y_{o},x_{o})$ to $(y'_{o},x'_{o})$ with
 \be
    y'_{o}&:=&y_{o}+\ell^{\eps}(x_{o}-x'_{o})\;,\label{eq: almost opt strat outside y}\\
    x'_{o}&:=& x_{o}+[-\eps\; \check \xib(t_{o},s_{o})+\theta(t_{o},s_{o})-x_{o}]^{+} - [x_{o}-\eps \;\check \xib(t_{o},s_{o})-\theta(t_{o},s_{o})]^{+}.~~~\label{eq: almost opt strat outside x}
 \ee
By  Remark \ref{rem: encadrement v eps}, one has
\b*
v^{\eps}(t_{o},s_{o},{p_{o},x_{o}})&\le& v^{\eps}(t_{o},s_{o},{p_{o},x'_{o}})+x'_{o}-x_{o}+\eps^{3}|x_{o}-x'_{o}| \\
&\le&  v^{\eps}(t_{o},s_{o},{p_{o},x'_{o}})+x'_{o}-x_{o}+\eps^{3} (C_{K}+|x_{o}|)\;,
\e*
in which the last inequality follows from Assumption \ref{ass: BS expo case}. Hence,
\b*
(v^{\eps}-v)(t_{o},s_{o},{p_{o},x_{o}})&\le&(v^{\eps}-v)(t_{o},s_{o},{p_{o},x'_{o}})+  x_{o}-x'_{o}+x'_{o}-x_{o}+ \eps^{3} (C_{K}+|x_{o}|)\\
&\le& (v^{\eps}-v)(t_{o},s_{o},{p_{o},x'_{o}})+ \eps^{3} (C_{K}+|x_{o}|).
\e*
Since $(t_{o},s_{o},x'_{o})$ belongs to the closure of $\Jc_{\eps}$, we can apply the analysis of the preceding steps to conclude.
	\ep\\

A by-product of the above argument is the explicit construction of a  strategy $L^{\eps}$ which is $O(\eps^{2})$-optimal for the problem with transaction costs. The constant $C_{K}$ in the following proposition can be recovered in terms of the constant $K$ of Assumption \ref{ass: BS expo case}.

\begin{proposition}\label{prop: strat eps opti expo} Let the conditions of Proposition  \ref{prop: ass ok for expo BS} hold. Then, there exists a constant $C_{K}>0$ such that the following holds: Fix $(t_{o},s_{o},x_{o},p_{o})\in [0,T]\x (0,\infty) \x \R\x (-\infty,0)$, $\epsilon \in (0,1)$, let
\b*
y_{o}&:=&\psi^{\eps}(t_{o},s_{o},p_{o},x_{o})+\eps^{2}\left(C_{K}+1\right),
\e*
where $\psi^{\eps}$ is defined as in \reff{eq: def psi eps proof expo},
 $(y'_{o},x'_{o})$ be defined as in \reff{eq: almost opt strat outside y}-\reff{eq: almost opt strat outside x},
 $L^{\eps,k}$ be given by the solution of  \reff{eq: skorokhod problem}-\reff{eq: def Lepsk expo}-\reff{eq: def k} for the initial condition $(t_{o},s_{o},x'_{o},y'_{o})$, and
 $
 L^{\eps}:=L^{\eps,k}+x'_{o}-x_{o},
$
then
\be\label{eq: prop: strat eps opti expo}
\Esp{ \Psi(\Delta_{t_{o},s_{o},y_{o},x_{o}}^{\eps,L^{\eps}})}\ge  p_{o}
\;\mbox{ and }\; y_{o}=v^{\eps}(t_{o},s_{o},p_{o},x_{o})+O(\eps^{2}).
\ee
\end{proposition}

\proof  We first prove the left-hand side inequality of \reff{eq: prop: strat eps opti expo}. When $(t_{o},s_{o},x_{o})$ belongs to the closure of $\Jc_{\epsilon}$ defined in \reff{eq: def domain reflexion Skorohod problem}, then $(x'_{o},y'_{o})=(x_{o},y_{o})$ and this is an immediate by-product  of the construction made in the proof of Proposition  \ref{prop: ass ok for expo BS}.
The general case is treated as in Step 5 of the proof  of Proposition  \ref{prop: ass ok for expo BS}, observing that
$$
\psi^{\eps}(t_{o},s_{o},p_{o},x_{o})=\psi^{\eps}(t_{o},s_{o},p_{o},x'_{o})+x'_{o}-x_{o}+\eps^{3}|x_{o}-x'_{o}|=\psi^{\eps}(t_{o},s_{o},p_{o},x'_{o})-\ell^{\eps}(x_{o}-x'_{o}),
$$
by Proposition \ref{prop: pi explicit BS example} and \reff{eq: explicit form of varpi}.

To prove the right-hand side identity in \reff{eq: prop: strat eps opti expo}, it suffices to use Proposition  \ref{prop: ass ok for expo BS}  and to recall \reff{eq: explicit form of varpi}:
$$
(\psi^{\eps}-v^{\eps})(t_{o},s_{o},p_{o},x_{o})=(\psi^{\eps}-v)(t_{o},s_{o},p_{o},x_{o})+(v-v^{\eps})(t_{o},s_{o},p_{o},x_{o})=O(\eps^{2}).
$$
\ep
\\

Under an additional regularity conditions, one can  obtain a strategy which is optimal at the leading order $\eps^{2}$.

\begin{proposition}\label{prop: strat eps opti expo eps cube} Let the conditions of Proposition  \ref{prop: ass ok for expo BS} hold.  Assume further that
$
|s^{2}\delta_{ss}|\le K\;\mbox{ on } \D.
$
Then, there exists $C_{K}>0$ such that the following holds:  Fix $(t_{o},s_{o},x_{o},p_{o})\in [0,T]\x (0,\infty) \x \R\x (-\infty,0)$, $\epsilon \in (0,1)$, set
\b*
y_{o}&:=&(v+\eps^2\hat u+\eps^{4}\varpi\circ \xib_{\eps}) (t_{o},s_{o},p_{o},x_{o})+\eps^{3}\left(C_{K}+1\right),
\e*
let $(y'_{o},x'_{o})$ be defined as in \reff{eq: almost opt strat outside y}-\reff{eq: almost opt strat outside x} with $\hat \xib $ in place of $\check \xib $,
 $L^{\eps,k}$ be given by the solution of  \reff{eq: skorokhod problem}-\reff{eq: def Lepsk expo} for $\Jc_{\eps}$ defined with $\hat \xib $ in place of $\check \xib $ and for
$$
 k:=-\eta^{-1}\ln\left(p_{o}(e^{-\eta \eps^{3}}-1)/C_{K}\right)+1
$$
and the initial condition $(t_{o},s_{o},x'_{o},y'_{o})$,  and set
 $
 L^{\eps}:=L^{\eps,k}+x'_{o}-x_{o},
$
then
\b*
\Esp{ \Psi(\Delta_{t_{o},s_{o},y_{o},x_{o}}^{\eps,L^{\eps}})}\ge  p_{o}
\;\mbox{ and }\; y_{o}=v^{\eps}(t_{o},s_{o},p_{o},x_{o})+O(\eps^{3}).
\e*
\end{proposition}

\proof We only sketch the proof since it is a straightforward adaptation of the proof of Proposition \ref{prop: strat eps opti expo}, see also the proof  of Proposition \ref{prop: ass ok for powe BS} below.

We follow line by line the arguments of the proof of Proposition \ref{prop: strat eps opti expo} but with $\psi^{\eps}$ and $\Jc_{\eps}$ defined by
\b*
\psi^{\eps}&:=&v+\eps^2\hat u+\eps^{4}\varpi\circ \xib_{\eps},
\\
\Jc_{\eps}&:=&\{(t,s,x)\in[0,T]\x (0,\infty)\x \R: -\hat \xib(t,s) <\xib_{\eps}(t,s,x)<\hat \xib(t,s)\}.
 \e*
The fact that $\hat u$ is a classical solution of
   \reff{eq: 2nde corrector equation} while $\varpi$ solves \reff{eq: first corrector equation}
 implies that   the counterpart of   \reff{eq: def Eeps proof expo} is
 \b*
            E_{\eps}&=&
            \int_{t_{o}}^T R_{\eps}\left(\tau,S_{\tau},P^{\eps}_{\tau},X^{\eps}_{\tau} \right) d\tau\;,
 \e*
   where $R_\eps$ is given by \reff{eq : def R eps iota} for $\phi:=\hat u$ and $w:=\varpi$.
   Observe that \reff{eq: x bounded in Jceps and hat a CL lipsch} remains in force since neither $\hat u$ nor $\varpi$ depend on $p$
   and $s\hat u_s$ and $s\varpi_s\circ\xib_\eps\mathbf1_{\bar \Jc_\eps}$ are bounded.
   Under our additional assumptions, it is easy to check from the proof of Lemma \ref{lem: remainder estimate}, see \reff{eq : def R eps iota}, that  $|E_{\eps}|\le \eps C_{K}$:
   the additional assumption that $s^2\delta_{ss}$ is bounded allows to control the term $\Lc_S\varpi$
   in $R^\eps_2$ whereas the other terms are bounded by Assumption \ref{ass: BS expo case}.

\ep


\subsection{Power case}

We now provide the proof of Proposition \ref{prop: ass ok for powe BS}.  Since it is very close to the one of Proposition \ref{prop: ass ok for expo BS}, we focus on  the differences.

{\bf Proof of Proposition \ref{prop: ass ok for powe BS}.}
{ We only show that, for any   compact subset $B_{o}\subset (-\infty,0)$, there exists $c_{{o}},\eps_{{o}}>0$ such that
   \be\label{eq: uniforml x linear growth ueps power case}
     u^{\eps}(\zeta,x)\le c_{o}(1+\eps  |x| ) \; \;\mbox{ for all } (\zeta,x)\in [0,T]\x (0,\infty)\x B_{o}\x \R \;\mbox{ , } \;\eps\in (0,\eps_{o}].
   \ee
}
From now on, we fix a compact subset  $B_{o} \subset (-\infty,0)$. We also fix another compact set $B\subset (-\infty,0)$ such that $B_{o}\subset {\rm Int}(B)$,  and denote by $C_{B}>0$ a generic constant that depends at most on $B$, and that may change from line to line. It will be clear later on that $B$ can be chosen in terms of $B_{o}$.

{\sl Step 1.} We first deduce from   \reff{eq: explicit price power case} that, for $(t,p)\in [0,T]\x (-\infty,0)$ one has
  \be
\left\{    \begin{array}{c}
  \theta(t,p) =\frac{\lambda m(t)}{\sigma (1+\beta)}(-p)^{-\frac1\beta}\;,\;
    \delta(t,p)  =\theta(t,p)(\frac{\lambda}{\sigma (1+\beta)}-1)
    \;,\;
    \hat a(p)=\frac{\lambda \beta}{\beta+1} (-p) \;,\\
   h(t,p)  = \frac{\sigma^{2}}{2}\frac{\pi_{pp}}{(\pi_{p})^{2}}(t,p)\hat \xib(t,p)^2\;,\;
    \hat \xib(t,p)=\left(\frac32 \delta(t,p)^{2} \frac{(\pi_{p})^{2}}{\pi_{pp}}(t,p)\right)^{\frac13}\;.
    \end{array}
    \right. \label{eq: power BS case - explicit functions}
    \ee
Let $(\varpi,h)$ be defined as in Lemma \ref{lem: existence and property varpi} and note that $(\varpi(\cdot, \xi),h)$ depends only on $p$, for $\xi \in \R$. Let {$\hat u$} be the solution of \reff{eq: first corrector equation}. It is not difficult to deduce from \reff{eq: power BS case - explicit functions} that  one has
\be\label{eq: u power case}
f(t,p)=f(t,-1)(-p)^{-\frac1\beta} \; \;\mbox{ with $f(\cdot,-1)\in C^{\infty}_{b}([0,T])$},\;\mbox{ $f\in \{\theta,\delta, \hat \xib,h,{\hat u}\}$.}
\ee

We set
\b*
\psi^{\eps}&=&v+\eps^{2}  {\hat u}+\eps^{4}  \varpi\circ \xib_{\eps}\;\;\mbox{ , }\;\;
\hat a^{\eps}:=\frac{-\bar \sigma_{0}^{\top} D\psi^\eps}{\psi^{\eps}_p}\;,
\e*
and
   \be
                    \Jc_{\eps}&:=&\{(t,p,x)\in[0,T]\x(-\infty,0)\x \R: -\hat \xib(t,p) <\xib_{\eps}(t,p,x)<\hat \xib(t,p)\}.
\label{eq: def domain reflexion Skorohod problem power case}
                    \ee
We now fix $(t_{o},s_{o},x_{o})$ in the closure of $\Jc_{\eps}$, the general case being handled as in Step 5 of the proof of Proposition \ref{prop: ass ok for expo BS}. We let $p_{o}\in B$ and
\be\label{eq: choice yo power case}
y_{o}:=c+\psi^{\eps}(t_{o},s_{o},p_{o},x_{o})\;,
\ee
for some $c>0$ to be chosen later on.
We next   define $(Y^{\eps},X^{\eps},S,L^{\eps},P^{\eps})$ as in the proof of Proposition \ref{prop: ass ok for expo BS} but with $(\varpi,\hat \xib)$ in place of $(\check \varpi,\check \xib)$, namely
        \beq\label{eq: skorokhod problem power}
        \left\{\bal
        		&P^{\eps}=p_{o}+\int_{t_{o}}^\cdot \hat a^\eps\left(\tau,S_\tau,P^{\eps}_\tau,X^{\eps}_\tau \right)dW_\tau\;,\\
                &X^{\eps}
                    = x_{o}+ \int_{t_{o}}^\cdot X^{\eps}_{\tau}\frac{dS_{\tau}}{S_{\tau}} + \int_{t_{o}}^\cdot dL^{\eps+}_\tau-\int_{t_{o}}^\cdot dL^{\eps-}_{\tau}\;,\\
                &(\cdot,P^{\eps},X^{\eps})\in\Jc_\eps\quad dt\otimes d\P\mbox{-a.e. on }[t_{o},T]\;,\\
                &L^{\eps\pm}=\int_{t_{o}}^\cdot\chi_{\{(\tau,P^{\eps}_\tau,X^{\eps}_\tau)\in\partial\Jc_{\eps}^\pm\}}dL^{\eps\pm}_\tau\;,
        \eal\right.\eeq
        and
               \b*
         Y^{\eps}
                    = y_{o}- \int_{t_{o}}^\cdot (1+\eps^{3})dL^{\eps+}_\tau + \int_t^\cdot (1-\eps^{3})dL^{\eps-}_\tau\;,\;\;y_{o}:=c+\psi^{\eps}(t_{o},s_{o},p_{o},x_{o}).
             \e*
 We claim that a solution exists  and that, for all $q>0$, there exists $C_{B}^{q}>0$, which depends only on $B$ and $q$, such that
\be\label{eq: unif Lq bound P}
\sup_{\eps\in (0,1]}\Esp{\sup_{t\in [t_{0},T]}\left(|P^{\eps}_{t}|^{ q}+|P^{\eps}_{t}|^{-q}\right)}\le C_{B}^{q}.
\ee
This will be proved in Step {3} below.
 Since ${\hat u},\varpi\ge 0$,  {and $\hat u$ does not depend on $x$, }the same arguments as in Step 2 of the proof of Proposition \ref{prop: ass ok for expo BS} leads to
        \be\label{eq: power Y lower intermediary bound}
        Y^{\eps}+\ell^{\eps}(X^{\eps}) &\ge& c+\psi^{\eps}(\cdot,S,P^{\eps},0)  -\eps^{2}E_{\eps}(\cdot)\ge c+v(\cdot,S,P^{\eps},0)  -\eps^{2}E_{\eps} ,
        \ee
       where
            \b*
            E_{\eps}&:=& \eps |X^{\eps}|+
            \int_{t_{o}}^\cdot \left(  \frac{\sigma^{2}}{2}\frac{\pi_{pp}}{(\pi_{p})^{2}} \xib_{\eps}^{2} + \frac{\sigma^{2}}{2} \delta^2(  \varpi_{\xi\xi}\circ\xib_{\eps})+\Hc   {\hat u}+R_{\eps}  \right)(\tau,X^{\eps}_{\tau},P^{\eps}_{\tau}) d\tau\\
            &=& \eps |X^{\eps}|+ \int_{t_{o}}^\cdot  R_{\eps}(\tau,S_{\tau},P^{\eps}_{\tau},X^{\eps}_{\tau})  d\tau,
            \e*
            in which the second equality follows from the fact that  ${\hat u}$ and $\varpi$ solve \reff{eq: 2nde corrector equation} and \reff{eq: first corrector equation}  respectively, and $R_{\eps}$ is defined in \reff{eq : def R eps iota} for $\phi:={\hat u}$ and $w:=\varpi$.  Observe that all the functions in the definition of $R_{\eps}$ are powers of the $p$-variable multiplied, at least, by $\eps$.   Moreover, the definition of $X^{\eps}$ combined with \reff{eq: def domain reflexion Skorohod problem power case} and \reff{eq: power BS case - explicit functions} implies that $X^{\eps}$  is also controlled by a polynomial in $|P^{\eps}|$. Namely, we can find $q_{\beta},C_{\beta}>0$, which only depend on $\beta$, such that
 \begin{equation}\label{eq: power case control Gamma}
\eps^{-1}\int_{t_{o}}^t |R_{\eps}(\tau,S_{\tau},P^{\eps}_{\tau},X^{\eps}_{\tau})|d\tau+  |X^{\eps}_{t}|\le   \Gamma^{\eps}_{t}:= C_{\beta}\sup_{[t_{o},t]}(1+|P^{\eps}|^{-q_{\beta}}+|P^{\eps}|^{q_{\beta}})\;,\;\;t\in [t_{o},T].
 \end{equation}
 We now take $c=3\eps^{5/2}$. Since $v\ge -\kappa$, \reff{eq: power Y lower intermediary bound} implies
        \b*
        Y^{\eps}+\ell^{\eps}(X^{\eps}) &\ge& -\kappa +2\eps^{5/2}  +\eps^{5/2}(1-\eps^{1/2}\Gamma^{\eps}) .
        \e*
  Let $\tau_{\eps}$ be the first time such that $Y^{\eps}+\ell^{\eps}(X^{\eps})$ is equal to $\eps^{5/2}-\kappa$. We let $(\tilde Y^{\eps},\tilde X^{\eps})$ be defined by the strategy in which we follow $L^{\eps}$ on $[t_{o},\tau_{\eps}[$ and liquidate the position at  $\tau_{\eps}$, i.e.
  $$
  (\tilde Y^{\eps},\tilde X^{\eps})=(Y^{\eps},X^{\eps})\1_{[\![t_{o},\tau_{\eps}\wedge T[\![}+ (Y^{\eps}_{\tau_{\eps}\wedge T},\ell^{\eps}(X^{\eps}_{\tau_{\eps}\wedge T}))\1_{[\![\tau_{\eps}\wedge T,T]\!]}.
  $$

   Note that this strategy is admissible by construction. Set $A_{\eps}:=\{\eps^{\frac12} \Gamma^{\eps}_{T}\le  1\}$.  The inclusion $A_{\eps}\subset \{\tau_{\eps}\ge T\}$ follows from the last inequality and the fact that $\Gamma^{\eps}$ is non-decreasing. We then obtain
 \be
 \Esp{\Psi(\tilde Y^{\eps}_{T}+\ell^{\eps}(\tilde X^{\eps}_{T}))}&\ge& \Esp{\Psi(2\eps^{5/2}+\Phi(P^{\eps}_{T}))\1_{A_{\eps}}}-|\Psi(\eps^{5/2}-\kappa)|\Pro{A^{c}_{\eps}}\nonumber\\
 &\ge&  \Esp{ P^{\eps}_{T}}-\left(\Esp{|P^{\eps}_{T}|^{2}}^{\frac12}+|\Psi(\eps^{5/2}-\kappa)|\right)\Pro{A^{c}_{\eps}}^{\frac12}\nonumber\\
 &= & p_{o}-\left(\Esp{|P^{\eps}_{T}|^{2}}^{\frac12}+|\Psi(\eps^{5/2}-\kappa)|\right)\Pro{A^{c}_{\eps}}^{\frac12}\;,
 \label{eq: EPsi lower bound power cas}
 \ee
 in which we used the fact that $P^{\eps}$ is a martingale by \reff{eq: unif Lq bound P}.
 We now appeal to  \reff{eq: unif Lq bound P}  and   \reff{eq: power case control Gamma}  to obtain
 $$
 \Esp{| P^{\eps}_{T}|^{2}}\le   C_{B} \;\mbox{ , }\;  |\Psi(\eps^{5/2}-\kappa)|= \frac1{\eps^{5\beta/2}}
\;\mbox{ and }\;
 \Pro{A^{c}_{\eps}}\le    \eps^{6+5\beta}\Esp{|\Gamma_{T}^{\eps}|^{12+10\beta}}\le  \eps^{5\beta} C_{B} \eps^{6}.
 $$
 Combining the above shows that, for some $c_{B}>0$, which only depends on $B$,
 $$
  \Esp{\Psi(\tilde Y^{\eps}_{T}+\ell^{\eps}(\tilde X^{\eps}_{T}))}\ge p_{o}-c_{B}\eps^{3},
 $$
 and therefore, by \reff{eq: choice yo power case}, our choice $c=3\eps^{5/2}$, the fact that  $\hat u, \varpi$ satisfy \reff{eq: u power case} and that $v^{\eps}$ is non-decreasing in $p$,
 \be\label{eq: almost done up to shift p power case}
 v^{\eps}(t_{o},s_{o},p_{o} -\tilde c_{B}\eps^{3},x_{o})\le v(t_{o},s_{o},p_{o},x_{o}) +\tilde c_{B}\eps^{5/2} ,
 \ee
 for some constant $\tilde c_{B}>0$ that only depends on $B$.

 {\sl Step 2.} Since  $\tilde c_{B}$ does not depend on $p_{o}\in B$, the above is true for any $p\in B$ in place of $p_{o}$.
Set   $\iota(p):=p+\eps^{5/2}$ for $p\in B_{o}$, recall that $B_{o}\subset {\rm Int}(B)$. Then,
$$
0>\iota(p) -\tilde c_{B}\eps^{3}=p +\eps^{5/2}-\tilde c_{B}\eps^{3}\ge p \;\;\mbox{ for all $p\in B_{o}$ and $0<\eps\le \eps_{B}$, }
$$
 for some $\eps_{B}\in (0,1)$ such that $p+\eps_{B}^{5/2}\in B$ for all $p\in B_{o}$. For the rest of the proof, we assume that $p_{o}\in B_{o}$.  Then, \reff{eq: almost done up to shift p power case} applied to $\iota(p_{o})$ in place of $p_o$ and the fact that $v^{\eps}$ is non-decreasing in $p$ imply that
 \b*
 v^{\eps}(t_{o},s_{o},p_{o},x_{o})\le v^{\eps}(t_{o},s_{o},\iota(p_{o}) -\tilde c_{B}\eps^{3},x_{o}) \le v(t_{o},s_{o},\iota(p_{o})  ,x_{o}) +\tilde c_{B}\eps^{5/2}.
 \e*
We now use \reff{eq: explicit price power case} to obtain
  \b*
 v^{\eps}(t_{o},s_{o},p_{o},x_{o})  \le v(t_{o},s_{o}, p_{o}, x_{o})+\eps^{5/2} \beta^{-1}|m(t_{o})|\left|p_{o}+\eps_{B}^{5/2}\right|^{-\frac1\beta-1} +\tilde c_{B}\eps^{5/2}.
 \e*
This proves \reff{eq: uniforml x linear growth ueps power case}.

 {\sl Step 3.} It remains to prove our claim. Using \reff{eq: u power case} and \reff{eq: explicit form of varpi} {below}, we obtain that {$\hat a^{\eps}$ is locally Lispchitz on $\Jc_{\eps}$} and {that} there exists a function $f\in C^{\infty}_{b}([0,T])$ such that
{$$
|\hat a^{\eps}(t,p,x)|\le\frac{(-p)^{\frac1\beta +1}}{m(t)+\eps^{2}f(t)}\left|-\sigma x+\eps^{4}\sigma x \varpi_{\xi}\circ \xi_{\eps}(t,p,x)\right|
 ,\;(t,p,x)\in [0,T]\x(-\infty)\x \R.
$$}
{It} follows from \reff{eq: u power case} and {(ii) of } Lemma \ref{lem: existence and property varpi} that
\be\label{eq: estimate hat a eps power case for skorohod}
|\hat a^{\eps}(t,p,x)|\le C_{K} |p|\;\;\;\mbox{ for $(t,p,x)\in \Jc_{\eps}$, }
\ee
{and for $\eps$ small enough with respect to $f$ and $m$. }
In particular, the existence to the system \reff{eq: skorokhod problem power} will automatically imply \reff{eq: unif Lq bound P}. For $\rho>0$, set $B_{\rho}:=[-e^{\rho},-e^{-\rho}]$ and let $\hat a^{\eps,\rho}$ be a Lipschitz function such that $\hat a^{\eps,\rho}=\hat a^{\eps}$ on $[0,T+1]\x B_{\rho}\x \R$ and $\hat a^{\eps,\rho}=0$ on $[0,T+1]\x B_{2\rho}^{c}\x \R$. Here all functions are extended to $[0,T+1]$ by taking their values at $T$ on $[T,T+1]$. The set $\Jc_{\eps}^{\rho}:=([0,T+1]\x (B_{2\rho})^{c}\x \R)\cap \Jc_{\eps}$ is bounded and it follows from \cite{dupuis1993sdes} that there exists a strong solution  $(P^{\eps,\rho},X^{\eps,\rho})$ to \reff{eq: skorokhod problem power} with $\hat a^{\eps,\rho}$ in place of $\hat a^{\eps}$. Let $\tau_{\eps}^{\rho}$ be the first time after $t_{o}$ when   $P^{\eps,\rho}$ reaches the boundary of $B_{\rho}$. For $\rho>|\ln(-p_{o})|$, $(X^{\eps,\rho},P^{\eps,\rho})$ solves \reff{eq: skorokhod problem power} on $[\![t_{o},\tau_{\eps}^{\rho}\wedge T]\!]$.
It follows from \reff{eq: estimate hat a eps power case for skorohod} that $\tau_{\eps}^{\rho}\wedge (T+1)$ converges to $T+1$ in probability as $\rho\to \infty$. Hence, after possibly passing to a subsequence $(\tau_{\eps}^{\rho_{n}})_{n\ge 1}$, it converges almost surely to $T+1$ as $n\to \infty$.
 Let us set
$$
(X^{\eps},P^{\eps}):=(x_{o},p_{o})\1_{\{t_{o}\}}+\sum_{n\ge 1}\1_{ ]\!]\tau_{\eps}^{\rho_{n-1}}\wedge T,\tau_{\eps}^{\rho_{n}}\wedge T]\!] }(X^{\eps,\rho_{n}},P^{\eps,\rho_{n}})
$$
with the convention $\tau_{\eps}^{\rho_{0}}:=t_{0}$. Since  $(X^{\eps,\rho_{n}},P^{\eps,\rho_{n}})$ $=$ $(X^{\eps,\rho_{n+k}},P^{\eps,\rho_{n+k}})$ on $[\![t_{o},\tau_{\eps}^{\rho_{n}}\wedge T]\!]$, for all $k\ge 1$, it  solves \reff{eq: skorokhod problem power} on each $[\![t_{o},\tau_{\eps}^{\rho_{n}}\wedge T]\!]$, $n\ge 1$.  Since $(\tau_{\eps}^{\rho_{n}}\wedge (T+1))_{n\ge 1}$ converges almost surely  to $T+1$ as $n\to \infty$,    $(X^{\eps},P^{\eps})$ solves \reff{eq: skorokhod problem power} on $[t_{o},T]$.
 \ep

 \begin{remark}\label{rem: strat eps opti power} The same arguments as in the proof of Propositions \ref{prop: strat eps opti expo} and \ref{prop: strat eps opti expo eps cube} show that the above allows to construct a strategy $L^{\eps}$, based on the sole knowledge of $v$, $\hat u$, $\varpi$ and $\theta$,  satisfying $\Esp{ \Psi(\Delta_{t_{o},s_{o},y_{o},x_{o}}^{\eps,L^{\eps}})}\ge  p_{o}$ for
 \b*
 y_{o}&=& (v+\eps^{2}  {\hat u}+\eps^{4}  \varpi\circ \xib_{\eps})(t_{o},s_{o},p_{o},x_{o}) + C\eps^{5/2}
 \\
 &=&v^{\eps}(t_{o},s_{o},p_{o},x_{o})+o(\eps^{2}),
 \e*
 where $C>0$ can be computed explicitly.
\end{remark}

\section*{Appendix}

We provide here the proofs of Theorem \ref{thm: pde veps}, Theorem  \ref{thm: pde v} and Proposition \ref{prop: explicit formula pi} for completeness. These results are essentially known but our framework requires some slight adjustments.
 \\

\noindent{\bf Proof of Theorems \ref{thm: pde veps} and \ref{thm: pde v}:} We focus on the proof of Theorem \ref{thm: pde veps}. Theorem \ref{thm: pde v} is proved by combining the following arguments with the results of \cite{bet09} instead of  \cite{BoDa10}.   The arguments of \cite{BoDa10} can not be applied per-se to obtain Theorem \ref{thm: pde veps} because their Standing Assumption 4 may not hold in our context. We explain briefly how to modify it. First, this does not alter the proof of (GDP1) in {\cite[Corollary 2.9]{BoDa10}}, which in turn leads to the viscosity supersolution property by the same arguments as in \cite[Section 5]{BoDa10}. Similarly, the proof of the subsolution property \cite[Section 5]{BoDa10} can be reproduced once   (GDP2) stated in  {\cite[Corollary 2.9]{BoDa10}} is valid. It is the case, by \cite{BoDa10}, if one imposes the additional constraints $ Y^{t,y,\eps,L} +\ell^{\eps}( X^{t,x,s,L})\ge -c$ on $[t,T]$, with $c>0$ fixed independent of the control $L$. Their standing Assumption 4 is then satisfied, see   \cite[Lemma 3.3]{Kabanov99} which imposes a uniform $L^{2}$ bound on the admissible controls $L$.   Then, the corresponding value function $v^{\eps,c}$ satisfies that its upper-semicontinuous envelope  $v^{\eps,c \;*}$ is a viscosity subsolution of   \reff{eq: dpe v eps} on $\{v^{\eps,c,\;*}(t,s,p,x)+\ell^{\eps}(x)> -c\}$, by  \cite[Section 5]{BoDa10}.  The sequence of corresponding operators converges to the one of \reff{eq: dpe v eps} as $c\to \infty$. By standard stability results for viscosity solutions, see e.g.~\cite{barles91}, this implies that the relaxed semi-limit $v^{\eps,\infty\;*}$ defined by
$v^{\eps,\infty\;*}(t,s,p,x):=\limsup_{(c,t',s',p',x')\to (\infty,t,s,p,x)} v^{\eps,c \;*}(t',s',p',x')$ is a viscosity subsolution of \reff{eq: dpe v eps}.   Note that $  v^{\eps,\infty\;*}\ge v^{\eps*}$ by monotonicity. It remains to check that the converse inequality holds. But the admissibility constraint entering in the definition of $ \Lmf^{\eps}$ means that, for all $\iota>0$, we can find $c_{\iota}>0$ such that $v^{\eps,c_{\iota}}\le v^{\eps}+\iota\le v^{\eps\;*}+\iota$.
\ep
\\

\noindent {\bf Proof of Proposition \ref{prop: explicit formula pi}:}   Let us first fix $z>\pi(t,s,p)$.  Then, we can find  $(\vc,\alpha)\in \Uc\x\Amf$ such that
                $
                    \Psi ( Z^{t,s,z,\vc}_T - g ( S^{t,s}_T  )  )\ge P^{t,p,\alpha}_T.
                $  Recall from the discussion after \reff{eq: def P} that we can restrict $P^{t,p,\alpha}_T$ to take values in the image of $\Psi$, and therefore in the domain of definition of $\Phi$.
                Since $\Phi$ is non decreasing, it follows that
                $
                   Z^{t,s,z,\vc}_T \ge g\left( S^{t,s}_T \right) + \Phi\left(P^{t,p,\alpha}_T\right).
                $
                Then,  the convexity of $\Phi$ and the fact that $\Phi'\circ I$ is the identity  imply
                \b*
                   Z^{t,s,z,\vc}_T &\ge& g\left( S^{t,s}_T \right) + \Phi\circ I(\hat q Q^{t,s}_{T})+ \Phi'\circ I(\hat q Q^{t,s}_{T})(P^{t,p,\alpha}_T- I(\hat q Q^{t,s}_{T}))
                   \\
                   &=& g\left( S^{t,s}_T \right) + \Phi\circ I(\hat q Q^{t,s}_{T})+ \hat q Q^{t,s}_{T} (P^{t,p,\alpha}_T- I(\hat q Q^{t,s}_{T})).
                \e*
            We conclude by taking expectation under $\Q^{t,s}$. Since   $Z^{t,s,z,\vc}$ is a $\Q^{t,s}$-supermartingale, as a local-martingale bounded from below, and   $P^{t,p,\alpha}$ a $\P$-martingale, the definition of $\hat q$ and $\Q^{t,s}$ lead to
            $y\ge \gamma+\hat q(p-p)=\gamma$, where $\gamma$  denotes the right-hand side term in \reff{eq: explicit formula pi in general case}. This shows that $\pi(t,s,p)\ge \gamma$.

            To see that the reverse inequality holds, just observe that our integrability condition imply that we can find $\vc_{n} \in \Uc{(t,s,z_{n})}$ such that
            $$
            Z^{t,s,z_{n},\vc_{n}}_{T}=H_{n}:=\left(g(S^{t,s}_{T})+ \Phi\circ I(\hat q Q^{t,s}_{T})\right)\vee (-n),
            $$
            in which $z_{n}:=\E^{\Q^{t,s}}[H_{n}]\downarrow \gamma$ as $n\to \infty$. Then,
            $
            \Esp{\Psi(H_{n}-g(S^{t,s}_{T}))}\downarrow \Esp{I(\hat q Q^{t,s}_{T})}=p,
            $
            by monotone convergence and definition of $\hat q$.\ep

\bibliographystyle{plain}

\end{document}